\documentclass[a4paper]{article}

 \pdfoutput=1
\usepackage{authblk}
\usepackage{xcolor}
\usepackage{comment}
\usepackage{upgreek}

\usepackage{savesym}
\usepackage{amsmath}
\numberwithin{equation}{section}
\usepackage{ccaption}
\usepackage{verbatim}
\usepackage{epsfig}
\usepackage{asymptote}
\graphicspath{{Figures/}} 
\usepackage{helvet}

\usepackage{mathabx}
\savesymbol{ulcorner}
\savesymbol{urcorner}
\savesymbol{llcorner}
\savesymbol{lrcorner}
\usepackage{amsfonts}
\restoresymbol{AMS}{ulcorner}
\restoresymbol{AMS}{urcorner}
\restoresymbol{AMS}{llcorner}
\restoresymbol{AMS}{lrcorner}
\usepackage[bbgreekl]{mathbbol}
\DeclareSymbolFontAlphabet{\mathbbm}{bbold}
\DeclareSymbolFontAlphabet{\mathbb}{AMSb}
\usepackage{amsthm}
\usepackage{cases}
\usepackage[T1]{fontenc}
\usepackage{amssymb}
\usepackage{algorithm}
\usepackage{algpseudocode}
\usepackage{soul}
\usepackage{mathtools}
\usepackage{graphicx}
\usepackage{pict2e}
\usepackage{mwe}
\usepackage{footmisc}
\usepackage[cal=boondox,scr=boondoxo]{mathalfa}
\usepackage{multirow}

\usepackage{todonotes}

\usepackage{varioref}
\usepackage{url}
\usepackage{chapterbib}
\usepackage{mwe}
\usepackage{hyperref}

\usepackage{atveryend}
\makeatletter
    \AfterLastShipout{\immediate\write\@mainaux{\relax}}%
\makeatother

\newif\ifHNNdraft
\HNNdrafttrue

\makeatletter

\let\original@writefile\@writefile

\def\@writefile #1#2#3{\def\HNN@fileext{#1}%
     \let\HNN@next\original@writefile
     \in@{#1}{toc,lof,lot}%
     \ifin@
       \ifx\newlabel#3%
          \def\HNN@next ##1##2\newlabel
         {\futurelet\HNN@tmp\HNN@writefile #2\empty\HNN@writefile}%
       \fi
     \fi
     \HNN@next {#1}{#2}#3
}%
\def\HNN@writefile
{%
    \ifx\HNN@tmp\contentsline\expandafter\HNN@writefile@a\else
                             \expandafter\HNN@writefile@abort
    \fi
}%
\def\HNN@writefile@a  #1#2#3\HNN@writefile #4%
{%
    \original@writefile{\HNN@fileext}{\contentsline{#1}{#2\HNN@showkey {#4}}#3}%
    \newlabel {#4}%
}%
\def\HNN@writefile@abort #1\HNN@writefile
{%
    \original@writefile{\HNN@fileext}{#1}\newlabel
}%
\protected\def\HNN@showkey #1%
         {\ifHNNdraft\space\space \textit{key:} \texttt{#1}\fi}

\ifHNNdraft

\fi
\makeatother

\usepackage{nomencl}
\makenomenclature

\newcommand{\materialPtManifold}{\mathcal{X}}

\newcommand{\barspJMat}{\boldsymbol{\bar{\tilde{\sf J}}}}

\renewcommand{\u}[1]{\boldsymbol{#1}}

\renewcommand{\t}[1]{\tilde{#1}}

\renewcommand{\c}[1]{\mathcal{#1}}
\newcommand{\lsc}[2][\mathscr{l}]{{}^{ #1 }\! #2}
\newcommand{\lscH}[1]{\lsc[H]{#1}}
\newcommand{\dsf}[1]{\Delta\boldsymbol{\sf #1}}
\newcommand{\tpsb}[1]{\left. #1 \right.^{\sf T}}
\newcommand{\tps}[1]{\left( #1 \right)^{\sf T}}
\newcommand{\usf}[1]{\u{\sf #1}}
\newcommand{\busf}[1]{\bar{\usf{ #1}}}
\newcommand{\tu}[1]{\tilde{\u{ #1}}}
\newcommand{\tusf}[1]{\tilde{\usf{ #1}}}
\newcommand{\btusf}[1]{\bar{\tusf{ #1}}}
\newcommand{\pr}[1]{\left( #1 \right)}

\definecolor{carmine}{rgb}{0.59, 0.0, 0.09}

\title{
	An accelerometer-only algorithm for determining the acceleration field of a rigid body, with application in studying the mechanics of mild Traumatic Brain Injury
}
\author[1]{Mohammad Masiur Rahaman}
\author[1]{Wenqiang Fang}
\author[2]{Alice Lux Fawzi}
\author[1]{Yang Wan}
\author[1,*]{Haneesh Kesari}
\affil[1]{Brown University School of Engineering, 184 Hope St., Providence, RI, USA}
\affil[2]{Institute for Molecular and Nanoscale Innovation, Brown University, 184 Hope St., Providence, RI, USA}
\affil[*]{Corresponding author, haneesh\_kesari@brown.edu}
\begin{document}
\maketitle
\begin{abstract}

We present an algorithm for determining the acceleration field of a rigid body using measurements from four tri-axial accelerometers.
The acceleration field is an important quantity in bio-mechanics problems, especially in the study of mild Traumatic Brain Injury (mTBI).
The \textit{in vivo} strains in the brain, which are hypothesized to closely correlate with brain injury,
are generally not directly accessible outside of a laboratory setting. However, they can be estimated on knowing the head's acceleration field.
In contrast to other techniques, the proposed algorithm uses data exclusively from accelerometers, rather than from a combination of accelerometers and gyroscopes.
For that reason, the proposed accelerometer only (AO) algorithm does not involve any numerical differentiation of data, which is known to greatly amplify measurement noise.
For applications where only the magnitude of the acceleration vector is of interest, the algorithm is straightforward, computationally efficient and does not require computation of angular velocity or orientation.
When both the magnitude and direction of acceleration are of interest, the proposed algorithm involves the calculation of the angular velocity and orientation as intermediate steps.
In addition to helping understand the mechanics of mTBI, the AO-algorithm may find widespread use in several bio-mechanical applications, gyroscope-free inertial navigation units, ballistic platform guidance, and platform control.


\end{abstract}


\section{
Introduction
}
\label{sec:Introduction}

Commonly referred to as a ``concussion,'' mild Traumatic Brain Injury (mTBI) occurs when the head is exposed to blunt trauma, which results in the brief alteration of a person's mental status (such as confusion or disorientation) \cite{national2003report,blyth2010traumatic}.
Clinical research has shown that  mTBI causes serious and lasting health problems with more than 1.5 million cases per year in the United States alone~\cite{national2003report,langlois2006traumatic, laker2011epidemiology}.
The change in mental status following mTBI may be brief, but the damage to neural tissue is permanent and accumulated damage from repetitive head injuries can lead to progressive neurodegenerative disease \cite{blyth2010traumatic,gavett2011chronic}.
The overarching objective from the bio-mechanics perspective in the direction of mTBI has been to gauge the potential of an impact type event in causing mTBI.

It has been hypothesized that the injury  takes place through the creation of large strains and strain-rates in the tissue~\cite{bar2016strain}.
Outside of a laboratory setting, \textit{in vivo} tissue strains are generally not directly accessible during injury inducing events (e.g., impacts).
For that reason, there is  interest in developing indirect strategies for estimating  the strains and strain-rates in the brain.
One such indirect strategy is as follows.

In blunt impact events that may lead to mTBI the strains in the skull are usually elastic and small.
Since the deformation induced in the brain due to such strains is unlikely to cause mTBI it is reasonable to ignore them.
Thus, the strains and strain-rates in the brain can be estimated by \textit{(i)} modeling the skull as a rigid body, \textit{(ii)} determining its accelerations in an impact type event, and then \textit{(iii)} using those accelerations to setup a continuum mechanics based calculation for the brain's deformation.
In this paper we focus on step \textit{(ii)} of this indirect strategy.

There, of course,  exist several alternate strategies that do not involve rigid body based models~\cite{bayly2005nonrbd,bain2000nonrbd,Chan2018nonrbd,gomez2019nonrbd}.
However, for several reasons rigid body based models continue to be invaluable tools in understanding the mechanics of mTBI.
Some of those reasons are related to the analytical and computational tractability of rigid body models, while still others are related to the rigid body models' amenability in interfacing with neck and body models~\cite{wright2013multiscale}.

It is possible to measure the accelerations of any finite number of points on the head by attaching accelerometers to a helmet or other gear so that they are in close contact with those points.
We assume that  the acceleration of a point on the skull correlates closely with the point that lies directly above it on the head and use the terms ``head'' and ``skull'' interchangeably.
For the purposes of carrying out step \textit{(iii)} such that the strains in the brain can be spatially resolved it is necessary to know the acceleration at all head points.
Therefore, in this paper we present an algorithm for constructing a time sequence of head acceleration fields from a finite number of accelerometer measurements.
The head acceleration field is a map that given a point on the head returns the acceleration at that point.
Since the algorithm that we present uses data only from accelerometers we refer to it as \textit{accelerometer only} (AO) algorithm.

In contrast, many of the popular algorithms used for estimating acceleration fields use data from other sensors, such as gyroscopes and magnetometers, in addition to those from accelerometers~\cite{camarillo2013instrumented,allison2015gyro,WU2016gyro}.

There exist a plethora of techniques for determining the acceleration field of a rigid body.
Many of those techniques, however, introduce approximations that are only justified if the rotations contained in the rigid body's motion are small (e.g., see ~\cite{franck2015extracting,crisco2004algorithm}).
The assumption of small rotations
was, perhaps, motivated by the belief that  during blunt impact events that may lead to mTBI  the rotational component of the head's motion is small, or  the peak strains inside the head primarily correlate with the non-rotational component of the motion.
In the development of the AO-algorithm that we present in this paper we do not assume the rotations to be small.

The majority of the techniques that do not assume the rotational component of the motion to be negligible are based on inertial navigation technology.
Inertial navigation technology is a well established field for determining the orientation and position of a rigid body using measurements from what are termed inertial measurement units (IMUs)~\cite{britting1971inertial,titterton2004strapdown}.
Strapdown IMUs are a special type of IMUs that are especially  suitable for  biomechanics applications owing to their low weight and footprint.
Typically, these units consist of three gyroscopes and one tri-axial accelerometer.
However, knowing the  orientation and position in space does not immediately yield information about the acceleration field of the rigid body.
Determining the acceleration field requires taking two time derivatives of the orientation and position information.
Thus, determining  accelerations from strapdown IMUs requires taking at least one time derivative of the measured time signals.
It is well know that differentiating data numerically greatly amplifies its noise~\cite{ovaska1998noise,alonso2005noise}.
The AO-algorithm that we present does not involve any numerical differentiation of data.
It does, however, involve numerical integrations.
Thus, potentially, the AO-algorithm might be more sensitive to bias type errors than strapdown IMU based algorithms.

In addition to not involving numerical differentiation there are other potential benefits to the AO-algorithm  over those based on strapdown IMUs.
Recall that algorithms based on strapdown IMUs make use of gyroscopes, whereas the AO-algorithm exclusively makes use of accelerometers.
Gyroscopes can measure angular velocity directly, but MEMS gyroscopes have limited sensitivity.
Optical gyroscopes possess remarkable sensitivity, but generally they are too large for biomechanics and other related applications~\cite{khial2018nanophotonic}.
Commercially available accelerometers are inexpensive and excel in dynamic range, accuracy, resolution, and response time~\cite{zou2018algorithm,tan2005design} making accelerometer-only systems appealing as a means to measure the acceleration of a rigid body.

Techniques that determine the acceleration field exclusively from accelerometers already exist.
However, the AO-algorithm is more general than those techniques.
For example, compared to the technique presented by Padagaonkar \textit{et al.} \cite{padgaonkar1975measurement,zou2014isotropic} the location and orientation of the accelerometers in the AO-algorithm can be fairly arbitrary.
The AO-algorithm can be applied to 3D motions whereas the one by Cardou \cite{cardou2008estimating} can only be applied to 2D motions.

In \S\ref{Sec:SOA} we present the mathematical formulation of rigid body motion that is required for the development of the AO-algorithm.
The theory of rigid body motion (RBM) is a core topic in the subject of classical mechanics~\cite{goldstein2002classical,landau1978course}.
For a more mathematical treatment of RBM from a perspective of geometric mechanics, please see~\cite[specifically, ch. 9]{maruskin2012introduction}~\cite{marsden2013introduction}.
For a modern, continuum mechanics style treatment of RBM, please see the excellent works of Jog~\cite[especially chs. 2, 3, and 8]{jog2015continuum} and O'Reilly~\cite{o2008intermediate}.

Our formulation contains some new and interesting features.
We found these features to be necessary for a mathematically rigorous and consistent development of the AO-algorithm. Some novel features of our formulation are as follows: \textit{(a.)\hypertarget{text:comm_a}} The rigid body is modeled as a topological space.
This allows for the  consideration of more general types of solids.
For details see the beginning paragraphs of \S\ref{sec:GeneralBalanceLawDerivation}.
\textit{(b.)} The forces are modeled as measures, which allows for a consistent treatment of point forces.
This generalized way of modeling forces in RBM can also be found in~\cite{oliva2004geometric}.
\textit{(c.)} A notable feature of our formulation is that positions, velocities, and accelerations belong to different vector spaces.
This feature makes our formulation more consistent with geometric mechanics based formulations than classical formulations.
In classical formulations there are no notions of tangent spaces.
In our formulation the velocity vector space can be identified with the trivial tangent bundle of the Euclidean point space in which the rigid body's material points execute their motion.
Another consequence of this feature is that in our formulation the physical notion of units finds a perfect home in the mathematical notion of vector spaces.
In  classical treatments, all positions, velocities, and accelerations share the same basis vectors, with the information about their units carried by the components of the basis vectors.
This aspect of the classical treatments makes them inconsistent with the modern mathematical theory of vector spaces, where the components are required to belong to a field, typically real  or complex numbers, which is generally shared by all the vector spaces involved in the physical theory, and hence do not have  units.
In our formulation,  positions, velocities, and accelerations all have different sets of basis vectors.
Since basis vectors are abstract entities, we can use them to store information about the units.
A vector's components in our formulation belong to the set of real numbers, and thus have no units.
In addition to our formulation being more mathematically consistent than  classical treatments, it also has the practical advantage that all variables and equations in it come out automatically non-dimensionalized.

Following \S\ref{Sec:SOA},  in \S\ref{Sec:PA} we develop the AO-algorithm.
In \S \ref{Sec:NumericaResults} we demonstrate  the predictive capability of the AO-algorithm using a numerical simulation of impact between a rigid body and an elastic half-space.
We make a few closing remarks in  \S\ref{sec:Discussions}.

\section{Rigid body motion}
\label{Sec:SOA}
\setcounter{equation}{0}

In this section we present a  formulation of RBM that is different from what is found in standard treatments of this subject (see our comments \hyperlink{text:comm_a}{\textit{(a)--(c)}} in \S\ref{sec:Introduction}). As we stated in \S\ref{sec:Introduction}, we found this formulation necessary for a mathematically consistent and rigorous development of the AO-algorithm. On a cursory level, our formulation of RBM might appear to be similar to that of standard continuum mechanics. However, there are important differences between the two, which we emphasize in \S\ref{sec:kinematics} and~\S\ref{sec:GeneralBalanceLawDerivation}.

\subsection{Kinematics}
\label{sec:kinematics}

Let $\mathbb{E}$ be a finite dimensional, oriented,  Hilbert space\footnote{A Hilbert space is any complete, inner-product space. Some authors, e.g. Kolmogorov and Fomin~\cite{kolmogorov100introductory}, reserve the term ``Hilbert space'' for referring to only infinite dimensional, complete, inner-product spaces, and refer to finite dimensional (complete) inner-product spaces as ``Euclidean spaces.'' However, we choose to use the term ``Hilbert spaces'' to refer to finite dimensional (complete) inner-product spaces as well since we believe that the use of the term ``Euclidean spaces'' would contribute to suppressing the abstract nature of the velocity and the acceleration spaces, $\mathbb{V}$ and $\mathbb{A}$, respectively.  We only consider finite dimensional spaces in our work. Since all finite dimensional inner-product spaces are complete, instead of ``Hilbert space'' we could have equivalently also used the term ``inner-product space.''}. The point space $\mathcal{E}$ is $\mathbb{E}$'s principal homogeneous space. It is possible to establish a one-to-one correspondence between the points in $\mathcal{E}$ and the vectors in $\mathbb{E}$. For example, choosing a point $o\in \mathcal{E}$, which we call $\mathcal{E}$'s origin,  we say that the point $x\in \mathcal{E}$ and the vector  $\boldsymbol{x}\in \mathbb{E}$ correspond to each other iff $o+\boldsymbol{x}=x$. We consider the rigid solid $\mathcal{B}$ to
be a topological space\footnote{cf.~\cite[p. 37]{truesdell2004non}, where, as is standard in continuum mechanics, the body $\mathcal{B}$ is modeled as a manifold.}. The motion of the rigid solid takes place in $\mathcal{E}$. Therefore, we formally define the vectors in $\mathbb{E}$ to have units of length, say meters, and  refer to $\mathbb{E}$ as the physical-space. Let $\mathbb{E}_{\rm R}$ be another  oriented Hilbert space, distinct from $\mathbb{E}$, and let the point space $\mathcal{E}_{\rm R}$ be its principle homogeneous space. The space $\mathbb{E}_{\rm R}$ has the same dimension as $\mathbb{E}$.  We take the vectors in $\mathbb{E}_{\rm R}$ to have dimensions of length as well, and call $\mathbb{E}_{\rm R}$ the reference space.

\begin{figure}[ht!]
  \centering
\includegraphics[width=\textwidth]{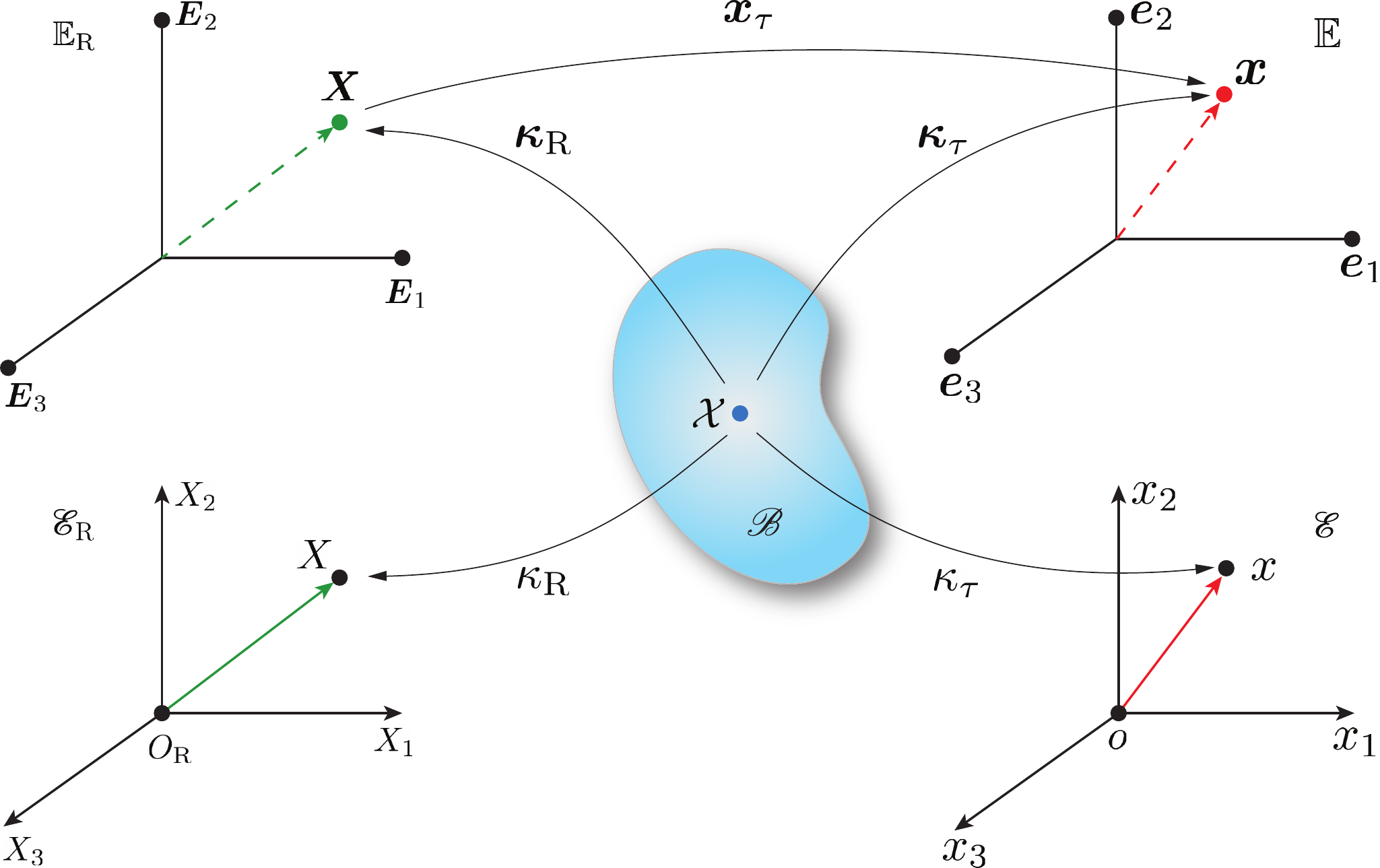}
\caption{The mathematical spaces related to the kinematics of rigid body motion. See \S
\ref{sec:kinematics} for details. 
}
\label{fig:VectorSpace}
\end{figure}

We designate a  select continuous, injective map from $\mathcal{B}$ into $\mathbb{E}_{\rm R}$ as the reference configuration  $\boldsymbol{\kappa}_{\rm R}$.
We call $\boldsymbol{\kappa}_{\rm R}(\mathcal{X})$ the particle $\mathcal{X}$'s reference position vector.
Selecting a point $O_{\rm R}\in\mathcal{E}_{\rm R}$ as $\mathcal{E}_{\rm R}$'s origin, to $\boldsymbol{\kappa}_{\rm R}$ we associate the map $\kappa_{\rm R}:\mathcal{B}\to\mathcal{E}_{\rm R}$ such that $\kappa_{\rm R}(\mathcal{X})=O_{\rm R}+\boldsymbol{\kappa}_{\rm R}(\mathcal{X})$. When we refer to a vector $\boldsymbol{X}$ belonging to  $\boldsymbol{\kappa}_{\rm R}(\mathcal{B})$ as a material particle we in fact mean the material particle $\boldsymbol{\kappa}_{\rm R}^{-1}(\boldsymbol{X})$,  which belongs to $\mathcal{B}$. We call a continuous, injective map from $\mathcal{B}$ into $\mathbb{E}$ a \textit{configuration}, and the family of all configurations the configuration manifold $\mathfrak{C}$. To each $\boldsymbol{\kappa}\in \mathfrak{C}$ we  associate the map $\kappa:\mathcal{B}\to \mathcal{E}$ defined such that $\kappa(\mathcal{X})=o+\boldsymbol{\kappa}(\mathcal{X})$ (see Fig.~\ref{fig:VectorSpace}).

We denote the inner product between vectors $\boldsymbol{u}_i$ and $\boldsymbol{u}_j$ that belong to the Hilbert space $\mathbb{U}$ as $\boldsymbol{u}_i\cdot_\mathbb{U}\boldsymbol{u}_j$. We denote the norm that is induced by the inner product on $\mathbb{U}$ of $\boldsymbol{u}_i$ i.e., $\left(\boldsymbol{u}_i\cdot_\mathbb{U}\boldsymbol{u}_i\right)^{1/2}$ as $\left\Vert\boldsymbol{u}_i\right\Vert_\mathbb{U}$. However, when there is no risk of confusion, we will generally omit the subscripts of the $\cdot$ symbol and the $\left\Vert\cdot\right\Vert$ operator. We take the sets $\left(\boldsymbol{E}_i\right)_{i\in\mathcal{I}}$ and $\left(\boldsymbol{e}_i\right)_{i\in\mathcal{I}}$, where $\mathcal{I} =\left(1,\ldots, n_{\rm sd}\right)$, to form bases for $\mathbb{E}_{\rm R}$ and $\mathbb{E}$, respectively.
Here $n_{\rm sd}$ is the dimension of $\mathbb{E}_{\rm R}$ or $\mathbb{E}$.
The sets $\left(\boldsymbol{E}_i\right)_{i\in\mathcal{I}}$ and $\left(\boldsymbol{e}_i\right)_{i\in\mathcal{I}}$ are orthonormal, by which we mean $\boldsymbol{E}_i\cdot_{\mathbb{E}_{\rm R}}\boldsymbol{E}_j = \delta_{ij}$ and $\boldsymbol{e}_i\cdot_\mathbb{E}\boldsymbol{e}_j = \delta_{ij}$, where $i,j\in\mathcal{I}$. Here $\delta_{ij}$ is the Kronecker delta symbol, which equals unity iff $i=j$ and vanishes otherwise.

For $m$,~$n\in \mathbb{N}$, the set of natural numbers, we denote the space of all $m\times n$ real matrices as $\mathcal{M}_{m,n}(\mathbb{R})$ and identify $\mathcal{M}_{m,1}(\mathbb{R})$ with $\mathbb{R}^m$. When the vector $\boldsymbol{X}$ is represented as  $\sum_{i\in\mathcal{I}} X_i \boldsymbol{E}_i$, where $X_i\in\mathbb{R}$, we call $X_i$ the components of $\boldsymbol{X}$ w.r.t $(\boldsymbol{E})_{i\in \mathcal{I}}$ and $\left(X_i\right)_{i\in\mathcal{I}}\in\mathcal{M}_{n_{\rm sd},1}(\mathbb{R})$ the matrix representation of $\boldsymbol{X}$ w.r.t $\left(\boldsymbol{E}_i\right)_{i\in\mathcal{I}}$.  Similarly, when $\boldsymbol{x} = \sum_{i\in\mathcal{I}} x_i \boldsymbol{e}_i$, where $x_i\in \mathbb{R}$, we call $\left(x_i\right)_{i\in\mathcal{I}}\in\mathcal{M}_{n_{\rm sd},1}(\mathbb{R})$  the matrix representation of $\boldsymbol{x}$ with respect to $\left(\boldsymbol{e}_i\right)_{i\in\mathcal{I}}$. From here on, unless otherwise specified, we will be following the Einstein summation convention, for which a repeated index in a term will imply a summation of that term with the repeated index taking values in $\mathcal{I}$. Hence, we will be writing expressions such as $\sum_{i\in\mathcal{I}} X_i \boldsymbol{E}_i$ simply as $X_i \boldsymbol{E}_i$. In the following sections there are, however, terms in which a repeated index does not imply a sum. We will identify such terms using a parenthetical remark. When an index appears only once in a term, then unless otherwise specified, that term is understood to represent a tuple of terms with the unrepeated index ranging over $\mathcal{I}$.

We model time as a one-dimensional normed vector space $\mathbb{T}$ \footnote{Typically, time is  modeled as an element of $\mathbb{R}_{\ge 0}$. The choice of the positive real number line to model time stems from the fact that  the observation of the mechanical system begins at some fixed time instance and there is interest in  observing the mechanical system at only those time instances that come after the fixed time instance.

We would like the difference between two time instances to also be a valid time instance. That is, we would like time to be a vector space. This is important, e.g., for giving a rigorous definition to the velocity of a material particle that is, e.g., executing its motion in $\mathbb{E}$. The set $\mathbb{R}_{\ge 0}$, interpreted in the standard way, is, of course, not a vector space. For that reason, we believe that the entire real number line is a better model for time. From the modeling perspective there are no issues in modeling time as the entire real number line, with the tacit assumption that the values of the different physical entities are only meaningful at time instances that come after the fixed time instance.

Furthermore, keeping in line with our spirit of retaining the distinction between different physical vectors spaces even when they have the same dimension\footref{note2} we model time as a one dimensional normed vector space $\mathbb{T}$, which is distinct from $\mathbb{R}$. We choose the null vector in $\mathbb{T}$ to be the fixed time instance at which the mechanical system's observation begins.}.  We denote a typical point in $\mathbb{T}$ as $\boldsymbol{\tau}=\tau\boldsymbol{s}$, where $\tau\in \mathbb{R}$ and $\boldsymbol{s}$ is a fixed vector in $\mathbb{T}$ of unit norm. When we refer to ``the time instance $\tau$,'' ``the time $\tau$,'' or simply ``$\tau$'' we in fact mean the time point $\boldsymbol{\tau}$.

The solid's motion is a twice continuously differentiable map
\begin{align*}
&\mathbb{T}\to \mathfrak{C},\\
&\boldsymbol{\tau} \mapsto \boldsymbol{\kappa}_{\tau}.
\end{align*}
We refer to $\boldsymbol{\kappa}_{\tau}(\mathcal{X})\in \mathbb{E}$ as the material particle $\mathcal{X}$'s position vector at the time instance $\tau$. It can be shown that for the case of a rigid solid
\begin{equation}
\boldsymbol{\kappa}_{\tau}= \boldsymbol{T}_{\tau}\circ \boldsymbol{Q}_{\tau}\circ \boldsymbol{\kappa}_{\rm R},
\label{eq:DefCompos}
\end{equation}
where $\boldsymbol{Q}_{\tau}$ is a proper (orientation preserving), linear isometry from $\mathbb{E}_{\rm R}$ onto $\mathbb{E}$, the map $\boldsymbol{T}_{\tau}:\mathbb{E}\to \mathbb{E}$ is defined by the equation $\boldsymbol{T}_{\tau}(\boldsymbol{x})=\boldsymbol{x}+\boldsymbol{c}(\boldsymbol{\tau})$, where $\boldsymbol{c}(\boldsymbol{\tau})$ is the value of a vector valued function from $\mathbb{T}$ into $\mathbb{E}$.

We call the map $\boldsymbol{x}_{\tau}:\mathbb{E}_{\rm R}\to \mathbb{E}$ defined as $\boldsymbol{x}_{\tau}(\boldsymbol{X})
=\boldsymbol{T}_{\tau}\circ \boldsymbol{Q}_{\tau}(\boldsymbol{X})$, i.e.

\begin{equation}
\boldsymbol{x}_{\tau}(\boldsymbol{X})
=\boldsymbol{Q}_{\tau}\boldsymbol{X}+\boldsymbol{c}(\boldsymbol{\tau}),
\label{eq:PosVec}
\end{equation}
the deformation map. The vector $\boldsymbol{c}(\boldsymbol{\tau})$ can be written as $c_{i}(\tau)\boldsymbol{e}_i$, where $c_{i}:\mathbb{R}\to \mathbb{R}$. Similarly, the map $\boldsymbol{Q}_{\tau}$ can be written as $Q_{ij}(\tau)\boldsymbol{e}_i\otimes \boldsymbol{E}_j$, where $Q_{ij}:\mathbb{R}\to \mathbb{R}$ and

\begin{subequations}
\begin{align}
Q_{ki}(\tau)Q_{kj}(\tau)&=\delta_{ij},\label{eq:OrthonormalityConditiona}
\intertext{or equivalently}
Q_{ik}(\tau)Q_{jk}(\tau)&=\delta_{ij},\label{eq:OrthonormalityConditionb}
\end{align}
\label{eq:OrthonormalityCondition}
\end{subequations}
for all $\tau\in \mathbb{R}$. The expression $\boldsymbol{e}_i\otimes \boldsymbol{E}_j$ denotes  a linear map from $\mathbb{E}_{\rm R}$ into $\mathbb{E}$ that is defined by the equation $\left(\boldsymbol{e}_i\otimes \boldsymbol{E}_j\right)\boldsymbol{X}=\boldsymbol{e}_{i}\left(\boldsymbol{E}_j\cdot_{\mathbb{E}_{\rm R}} \boldsymbol{X}\right)$. In general, for any two vectors $\boldsymbol{u}_1\in\mathbb{U}$ and $\boldsymbol{w}_1\in\mathbb{W}$, where $\mathbb{U}$ and $\mathbb{W}$ are two arbitrary Hilbert spaces, $\boldsymbol{u}_1\otimes\boldsymbol{w}_1$ denotes the linear mapping from $\mathbb{W}$ into $\mathbb{U}$ that is defined as $\left(\boldsymbol{u}_1\otimes\boldsymbol{w}_1\right)\boldsymbol{w}_2 = \boldsymbol{u}_1\left(\boldsymbol{w}_1\cdot_\mathbb{W} \boldsymbol{w}_2\right)$ for all $\boldsymbol{w}_2\in\mathbb{W}$.

\paragraph*{Velocities} 
\smallskip 

Say $\mathbb{U}$ and $\mathbb{W}$ are two arbitrary Hilbert spaces. We denote the set of  bounded linear operators from $\mathbb{U}$ into $\mathbb{W}$ as $B(\mathbb{U},\mathbb{W})$. The motion of the particle $\mathcal{X}$ is the map $\boldsymbol{x}_{\mathcal{X}}:\mathbb{T}\to \mathbb{E}$ defined such that $\boldsymbol{x}_{\mathcal{X}}(\boldsymbol{\tau})=\boldsymbol{x}_{\tau}(\boldsymbol{\kappa}_{\rm R}(\mathcal{X}))$. It can be shown that the velocity of a material particle executing its motion in $\mathbb{E}$ lies in the space $B(\mathbb{T},\mathbb{E})=: \mathbb{V}$\footnote{\label{note2} The time vector space $\mathbb{T}$ is isomorphic to the Hilbert space $\mathbb{R}$. The  Hilbert spaces $\mathbb{E}_{\rm R}$, $\mathbb{E}$, $\mathbb{V}$, $\mathbb{A}$ etc. are similarly isomorphic to each other. However, we make a substantial amount of effort to retain the distinction between these spaces. The reason for our doing that is as follows. By the definition of a vector space, the operation of addition between any two elements belonging to the same vector space is well defined. So, if we were not to retain the distinction between the different vector spaces and identify them all with, say, $\mathbb{R}^3$ then that would allow for meaningless operations, such as the addition of a velocity vector to a position vector. Typically, this issue is circumvented by assuming that the coefficients of the different physical vectors w.r.t, say, a basis of $\mathbb{R}^3$, carry with them different units. However, we chose not to follow that formalism. In our formalism, the information contained in the units is stored in the basis vectors rather than in the coefficients. There are several advantages to our formalism. One of which is that the equations governing the coefficients come out to be non-dimensionalized by default. Also, our formalism is consistent with the mathematical theory of real vector and Hilbert spaces~\cite{halmos19421958}\cite[ch. 2--3]{kreyszig1978introductory}, where the coefficients belong to $\mathbb{R}$, i.e., they do not have any units attached to them, and the basis vectors themselves are abstract entities, and therefore can be taken to have units. The spirit that governs our formalism is best expressed in the following quote by Halmos~\cite[pp. 13--14]{halmos19421958} \textit{``One might be tempted to say that from now on it would be silly to try to preserve an appearance of generality by talking of the general n-dimensional vector space, since we know that from the point of view of studying linear problems isomorphic vector spaces are indistinguishable, and we may as well always study $\mathbb{R}^n$...we shall ignore the theorem just proved and treat n-dimensional vector spaces as self-respecting entities ...reason for doing this...vector spaces...would lose a lot of their intuitive content if we were to transform them into $\mathbb{R}^n$''}}. To be precise, the velocity of the particle $\mathcal{X}$ at the time instance $\tau$ is the operator $\boldsymbol{V}_{\mathcal{X}}(\boldsymbol{\tau})\in \mathbb{V}$ such that

\[
\lim_{\lVert \Delta \boldsymbol{\tau} \rVert \to 0}
\frac{
\lVert
\boldsymbol{x}_{\mathcal{X}}(\boldsymbol{\tau}+\Delta \boldsymbol{\tau})-\boldsymbol{x}_{\mathcal{X}}(\boldsymbol{\tau})-\boldsymbol{V}_{\mathcal{X}}(\boldsymbol{\tau})\Delta \boldsymbol{\tau}
\rVert
}
{
\lVert \Delta \boldsymbol{\tau} \rVert
} = 0.
\]
That is, $\boldsymbol{V}_{\mathcal{X}}(\boldsymbol{\tau})$ is the (Fr\'echet) derivative of $\boldsymbol{x}_{\mathcal{X}}$ at the time instance $\boldsymbol{\tau}$.
Hence, we refer to $\mathbb{V}$ as the (physical) velocity space.
The vectors $\boldsymbol{v}_i\in \mathbb{V}$, where $i\in \mathcal{I}$,  defined such that $\boldsymbol{v}_i \boldsymbol{\tau}=\tau \boldsymbol{e}_i$ form an orthonormal basis for $\mathbb{V}$.

The material velocity field  $\boldsymbol{V}_{\tau}: \mathbb{E}_{\rm R}\to \mathbb{V}$ is defined such that  $\boldsymbol{V}_{\tau}(\boldsymbol{\kappa}_{\rm R}(\mathcal{X}))=\boldsymbol{V}_{\mathcal{X}}(\boldsymbol{\tau})$. Specifically, for the  case of rigid body motion

\begin{equation}
\boldsymbol{V}_{\tau}(\boldsymbol{X})=\boldsymbol{L}_{\tau}\boldsymbol{X}+\boldsymbol{c}'(\boldsymbol{\tau}).
\label{eq:MatVelField}
\end{equation}
In $\eqref{eq:MatVelField}$  the map $\boldsymbol{L}_{\tau}:\mathbb{E}_{\rm R}\to \mathbb{V}$ is defined as $\boldsymbol{L}_{\tau}:=Q'_{ij}(\tau)\boldsymbol{v}_i\otimes \boldsymbol{E}_j$ and $\boldsymbol{c}'(\boldsymbol{\tau})=c'_i(\tau)\boldsymbol{v}_i$, where $Q'_{ij}$ and $c'_i$ are the derivatives of $Q_{ij}$ and $c_i$, respectively. Using $\eqref{eq:PosVec}$ and $\eqref{eq:MatVelField}$, it can be shown  that the velocity of the material particle  located at the time instance $\tau$  at a point $o+\boldsymbol{x}$ where $\boldsymbol{x}\in \boldsymbol{x}_{\tau}\circ\boldsymbol{\kappa}_{\rm R}(\mathcal{B})\subset\mathbb{E}$  is

\begin{equation}
\boldsymbol{v}_{\tau}(\boldsymbol{x})=\boldsymbol{W}_{\tau}(\boldsymbol{x}-\boldsymbol{c}(\boldsymbol{\tau}))+\boldsymbol{c}'(\boldsymbol{\tau}).
\label{eq:SpatialVelocityField}
\end{equation}
In $\eqref{eq:SpatialVelocityField}$ the operator $\boldsymbol{W}_{\tau}:\mathbb{E}\to \mathbb{V}$ is defined as

\begin{align}
\boldsymbol{W}_{\tau}:=\boldsymbol{L}_{\tau}\circ \boldsymbol{Q}_{\tau}^{*},
\label{eq:SpatialAngularVelocityTensor}
\end{align}
where the operator $\boldsymbol{Q}_{\tau}^{*}$ is the Hilbert-adjoint of  $\boldsymbol{Q}_{\tau}$. The Hilbert-adjoint of an operator $\boldsymbol{T}:\mathbb{U}\to \mathbb{W}$ is the operator $\boldsymbol{T}^*:\mathbb{W}\to \mathbb{U}$ defined such that $\boldsymbol{u}\cdot_{\mathbb{U}} \boldsymbol{T}^*\boldsymbol{w}=\boldsymbol{w}\cdot_{\mathbb{W}} \boldsymbol{T}\boldsymbol{u}$ for all $\boldsymbol{u}$ and $\boldsymbol{w}$ in $\mathbb{U}$ and $\mathbb{W}$, respectively.

\paragraph*{Accelerations}
\smallskip

The acceleration of the material particle $\mathcal{X}$ at the time instance $\boldsymbol{\tau}$ equals the value of the  (Fr\'echet) derivative of the map $\boldsymbol{\tau}\mapsto
\boldsymbol{V}_{\mathcal{X}}(\boldsymbol{\tau})$ at the time instance $\boldsymbol{\tau}$. It can be shown that the acceleration of the material particle $\mathcal{X}$ that is executing  its motion in $\mathbb{E}$ lies in the space $B(\mathbb{T},\mathbb{V})=:\mathbb{A}$. We refer to $\mathbb{A}$ as the (physical) acceleration space. The vectors $\boldsymbol{a}_i\in \mathbb{A}$, where $i\in \mathcal{I}$,  defined such that $\boldsymbol{a}_i \boldsymbol{\tau}=\tau \boldsymbol{v}_i$ form an orthonormal basis for $\mathbb{A}$.


For the case of a rigid body, the acceleration of the material particle $\boldsymbol{X}\in \mathbb{E}_{\rm R}$ at the time instance $\tau$ can be shown to be equal to
\begin{align}
\boldsymbol{A}_{\tau}(\boldsymbol{X})&=\boldsymbol{M}_{\tau}\boldsymbol{X}+\boldsymbol{c}''(\boldsymbol{\tau}), \label{eq:MatAccField}
\intertext{where}
\boldsymbol{M}_{\tau}&:=Q''_{ij}(\tau)\boldsymbol{a}_i\otimes \boldsymbol{E}_j,\label{eq:MtauDef}
\intertext{and}
\boldsymbol{c}''(\boldsymbol{\tau})&=c''_i(\tau)\boldsymbol{a}_i.\label{eq:cpp}
\end{align}
Here, $Q''_{ij}$ and $c''_i$ denote the second derivatives of $Q_{ij}$ and $c_i$, respectively.

Using $\eqref{eq:PosVec}$ and $\eqref{eq:MatAccField}$, it can be shown  that the acceleration of the material particle located at $\boldsymbol{x}\in \mathbb{E}$ at the time instance $\tau$ equals

\begin{align}
\boldsymbol{a}_{\tau}(\boldsymbol{x})&=\boldsymbol{M}_{\tau}\circ \boldsymbol{Q}_{\tau}^{*}(\boldsymbol{x}-\boldsymbol{c}(\boldsymbol{\tau}))+\boldsymbol{c}''(\boldsymbol{\tau}).
\label{eq:SpatialAccField}
\end{align}

\begin{figure}
    \centering
    \includegraphics[width=0.85\textwidth]{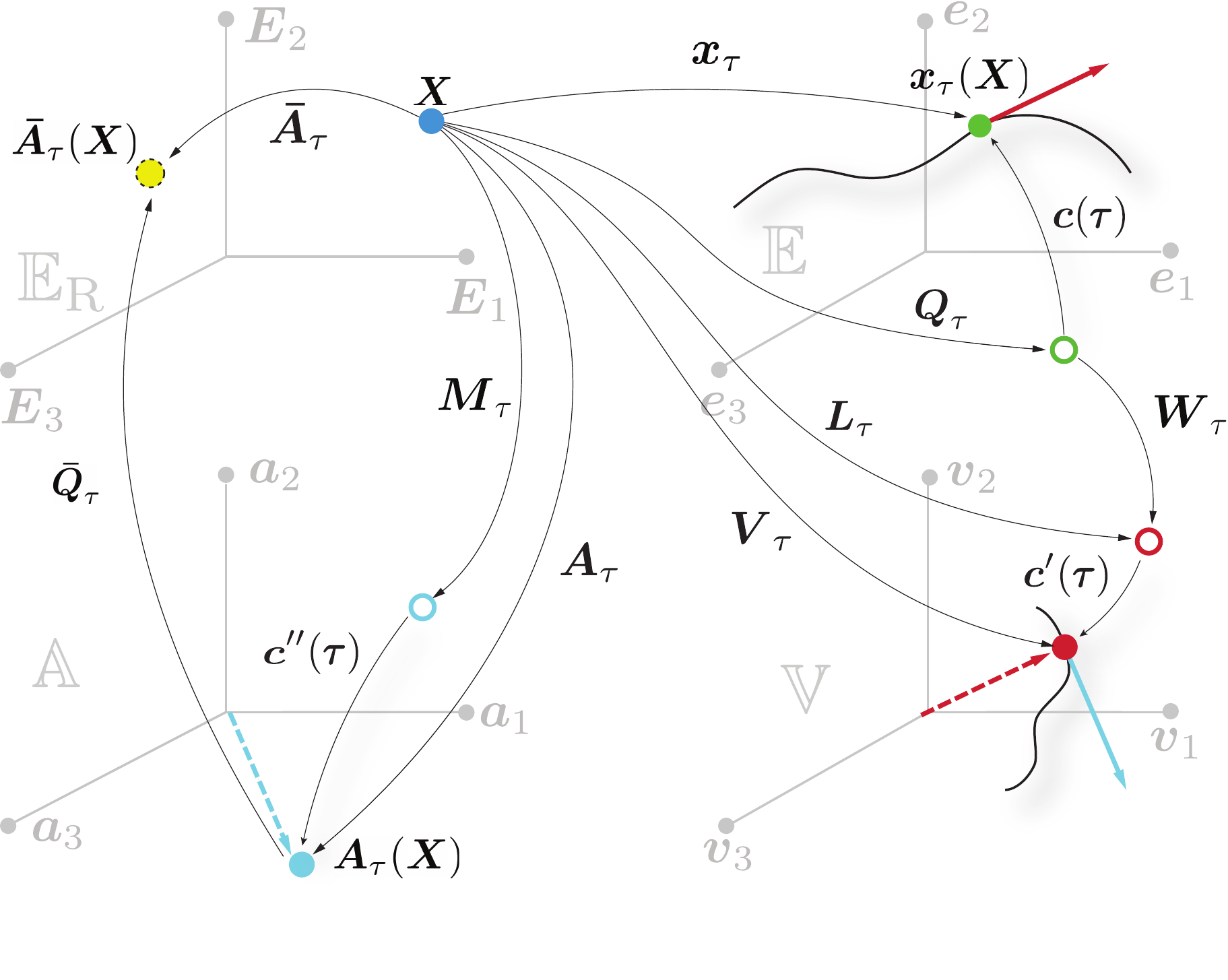}
    \caption{The mathematical spaces that appear in the kinematics and balance principles related to rigid body motion. See \S\ref{Sec:SOA} for details. 
		 The vectors $\boldsymbol{c}(\u{\tau})$, $\boldsymbol{c}'(\u{\tau})$ and $\boldsymbol{c}''(\u{\tau})$ belong to the spaces $\mathbb{E}$, $\mathbb{V}$, and $\mathbb{A}$, respectively. However, in the above figure they are to  be, respectively, interpreted as the maps
		$\mathbb{E}\ni \boldsymbol{x}\mapsto \boldsymbol{c}(\u{\tau})+\boldsymbol{x}\in\mathbb{E}$,
		$\mathbb{V}\ni \boldsymbol{v}\mapsto \boldsymbol{c}'(\u{\tau})+\boldsymbol{v}\in \mathbb{V}$,
		$\mathbb{A}\ni \boldsymbol{a}\mapsto \boldsymbol{c}''(\u{\tau})+\boldsymbol{a}\in \mathbb{A}$.
		The filled circle in $\mathbb{V}$, which also appears as the solid arrow in $\mathbb{E}$, denotes the velocity of the particle $\mathcal{X}$ at the time instance $\tau$. Similarly, the filled circle in $\mathbb{A}$, which also appears as the solid arrow in $\mathbb{V}$, denotes $\mathcal{X}$'s acceleration at the time instance $\tau$.
		}
    \label{fig:PseudoAcc}
\end{figure}

\paragraph*{Pseudo accelerations}
Both $\u{M}_{\tau}$ and $\u{c}''(\boldsymbol{\tau})$, are needed for computing the  acceleration of the particle $\boldsymbol{X}$ at the time instance $\tau$, $\boldsymbol{A}_\tau(\boldsymbol{X})$, from~\eqref{eq:MatAccField}. However, they are challenging to compute using the information provided by accelerometers. In \S\ref{Sec:PA} we will show that it is straightforward to use accelerometer information to compute the ``Pseudo-acceleration field'' $\bar{\u{A}}_{\tau}:\boldsymbol{\kappa}_{\rm R}\pr{\mathcal{B}}\to \mathbb{E}_{\rm R}$,
\begin{equation}
    \bar{\boldsymbol{A}}_\tau:= \bar{\boldsymbol{Q}}_\tau\circ\boldsymbol{A}_\tau,\label{eq:barAtauDef}
\end{equation}
where $\bar{\u{Q}}_{\tau}:\mathbb{A}\to \mathbb{E}_{\rm R}$ is defined by the equation
\begin{equation}
\bar{\boldsymbol{Q}}_\tau =Q_{ji}(\boldsymbol{\tau})\boldsymbol{E}_i\otimes\boldsymbol{a}_j.
\label{eq:Qbar}
\end{equation}
The quantity $\bar{\u{A}}_{\tau}(\u{X})$ is not $\u{X}$'s  acceleration. In fact, it does not even belong to $\mathbb{A}$, the acceleration space, in which $\u{X}$'s acceleration lies (see Fig.~\ref{fig:PseudoAcc}). However, interestingly, it can be shown that
\begin{equation}
\left\Vert\bar{\u{A}}_\tau(\boldsymbol{X})\right\Vert_{\mathbb{E}_{\rm R}}=
\left\Vert\boldsymbol{A}_\tau(\boldsymbol{X})\right\Vert_\mathbb{A},
\label{eq:NormEqual}
\end{equation} the magnitude of $\boldsymbol{X}$'s acceleration at the time instance $\tau$.  In  \S\ref{Sec:PA} we also present an algorithm for computing $\u{Q}_{\tau}$ from $\bar{\u{A}}_{\tau}$. Once $\u{Q}_{\tau}$ is known it can be used to construct $\bar{\u{Q}}_{\tau}$, using~\eqref{eq:Qbar}. The map $\bar{\u{Q}}_{\tau}$ can then be used to determine $\u{A}_{\tau}$, through~\eqref{eq:barAtauDef}, without the need to take  any derivatives of  $\u{Q}_{\tau}$. This is desirable since numerical differentiation of functions constructed using measured data can significantly add to the measurements' inherent error.

In \S\ref{Sec:PA} we present the procedure to compute $\bar{\u{A}}_{\tau}$ from experimental data. In order to do this, we operate on both sides of \eqref{eq:MatAccField} with $\bar{\boldsymbol{Q}}_\tau$ to find that
\begin{equation}
\bar{\boldsymbol{A}}_\tau(\boldsymbol{X}) = \boldsymbol{P}_\tau\boldsymbol{X} + \boldsymbol{q}(\boldsymbol{\tau}),
\label{eq:BarAcc}
\end{equation}
where
\begin{align}
\boldsymbol{P}_\tau &:= \bar{\boldsymbol{Q}}_\tau\circ\boldsymbol{M}_\tau,
\label{eq:PtauMap}
\end{align}
and  $\boldsymbol{q}:=\bar{\boldsymbol{Q}}_\tau\circ\boldsymbol{c}''$.

\subsubsection{Matrix representation}

We  sometimes abbreviate matrices using sans-serif bold face symbols. For example, we abbreviate the matrix $\left(Q_{ij}(\tau)\right)_{i,j\in\mathcal{I}}\in\mathcal{M}_{n_{\rm sd},n_{\rm sd}}(\mathbb{R})$ as $\boldsymbol{\sf Q}(\tau)$. In such cases we refer to the element in the   $i^{\rm th}$ row and $j^{\rm th}$ column of the matrix as $(\cdot)_{ij}$. That is, $\pr{\usf{Q}(\tau)}_{ij}=Q_{ij}(\tau)$.

\paragraph*{Positions and Orientations}


Expressing $\boldsymbol{Q}_{\tau}$, $\boldsymbol{X}$, $\boldsymbol{c}(\boldsymbol{\tau})$, and $\boldsymbol{x}_{\tau}(\boldsymbol{X})$ as $Q_{ij}(\tau)\boldsymbol{e}_i\otimes \boldsymbol{E}_j$, $X_i\boldsymbol{E}_i$,  $c_i(\tau) \boldsymbol{e}_i$, and $x_{\tau i}(\boldsymbol{X})\boldsymbol{e}_i$, respectively, in
$\eqref{eq:PosVec}$ we get that
\begin{equation}
x_{\tau i}(\boldsymbol{X})
=Q_{ij}(\tau)X_j+c_i(\tau).
\label{eq:PosVecCoeff}
\end{equation}
Defining $\usf{x}_{\tau}(\usf{X})=\left(x_{\tau i}(\boldsymbol{X})\right)_{i\in\mathcal{I}}$,  $\usf{c}(\tau)=\left(c_i(\tau)\right)_{i\in\mathcal{I}}$, and $\usf{I}=\left(\delta_{i j}\right)_{i,j\in\mathcal{I}}$ we can, respectively,  write $\eqref{eq:PosVecCoeff}$, and $\eqref{eq:OrthonormalityCondition}$ alternatively as
\begin{equation}
\usf{x}_{\tau}(\usf{X})
=\usf{Q}(\tau)\usf{X}+\usf{c}(\tau),
\label{eq:PosVecMatForm}
\end{equation}
and
\begin{subequations}
\begin{align}
\tpsb{\usf{Q}}(\tau)\,\usf{Q}(\tau)&=\usf{I},
\label{eq:OrthonormalityConditionMatrixForma}\\
\usf{Q}(\tau)\, \tpsb{\usf{Q}}(\tau)&=\usf{I},
\label{eq:OrthonormalityConditionMatrixFormb}
\end{align}
\label{eq:OrthonormalityConditionMatrixForm}
\end{subequations}
where $\tpsb{\usf{Q}}(\tau) = \tpsb{\left(\usf{Q}(\tau)\right)}$ and $\tpsb{\left(\cdot\right)}:\mathcal{M}_{m,n}(\mathbb{R}) \to  \mathcal{M}_{n,m}(\mathbb{R})$ is the transpose operator.
The matrix $\boldsymbol{\sf Q}(\tau)$, which we call the rotation matrix, belongs to the special orthogonal group $SO(n_{\rm sd})$.



\paragraph*{Velocities}
Expressing $\u{W}_{\tau}$ as $W_{ij}(\tau)\u{v}_i\otimes \u{e}_j$,
it can be shown using~\eqref{eq:SpatialAngularVelocityTensor} that
\begin{equation}
    \usf{W}(\tau)=\usf{Q}'(\tau)\tpsb{\usf{Q}}(\tau),
\label{eq:WMatDef}
\end{equation}
where
\begin{subequations}
\begin{align}
\boldsymbol{\mathsf{W}}(\tau)&:= \left(W_{ij}(\tau)\right)_{i,j\in\mathcal{I}},
\label{eq:ssW}
 \intertext{and}
W_{ij}(\tau)&:=Q'_{ik}(\tau)Q_{jk}(\tau).
\label{eq:SpatialAngularVelocityTensorCoeff}
\end{align}
\label{eq:WMatComp}
\end{subequations}
Substituting $\boldsymbol{W}_{\tau}$, $\boldsymbol{x}$, $\boldsymbol{c}(\boldsymbol{\tau})$, and $\boldsymbol{c}'(\boldsymbol{\tau})$ with $W_{ij}(\tau)\boldsymbol{v}_i\otimes \boldsymbol{e}_j$, $x_i\boldsymbol{e}_i$, $c_i(\tau) \boldsymbol{e}_i$, and $c'_i(\tau)\boldsymbol{v}_i$, respectively, in
$\eqref{eq:SpatialVelocityField}$ we can show that  $\boldsymbol{v}_{\tau}(\boldsymbol{x})$ can be expressed as $v_{\tau i}(\boldsymbol{x})\boldsymbol{v}_i$ where
\begin{equation}
v_{\tau i}(\boldsymbol{x})=W_{ij}(\tau)(x_j-c_j(\tau))+c'_i(\tau).
\label{eq:SpatialVelocityFieldCoeff}
\end{equation}
Defining $\boldsymbol{\mathsf{v}}_\tau(\boldsymbol{\mathsf{x}}) = \left(v_{\tau i}(\boldsymbol{x})\right)_{i\in\mathcal{I}}$, $\boldsymbol{\mathsf{x}}:= \left(x_i\right)_{i\in\mathcal{I}}$, $\boldsymbol{\mathsf{c}}(\tau)= \left(c_i(\tau)\right)_{i\in\mathcal{I}}$, and $\boldsymbol{\mathsf{c}}'(\tau)=\left(c'_i(\tau)\right)_{i\in\mathcal{I}}$ we can alternatively write $\eqref{eq:SpatialVelocityFieldCoeff}$ as
\begin{equation}
\boldsymbol{\mathsf{v}}_\tau(\boldsymbol{\mathsf{x}}) = \boldsymbol{\mathsf{W}}(\tau)(\boldsymbol{\mathsf{x}}-\boldsymbol{\mathsf{c}}(\tau))+\boldsymbol{\mathsf{c}}'(\tau).
\label{eq:SpatialVelocityFieldMatForm}
\end{equation}
It follows from $\eqref{eq:OrthonormalityConditionb}$ that $Q'_{ik}(\tau)Q_{jk}(\tau)= - Q_{ik}(\tau)Q'_{jk}(\tau)$,  which implies that $W_{ij}(\tau) = -W_{ji}(\tau)$. Hence, the matrix $\boldsymbol{\mathsf{W}}(\tau)$ belongs to the space of  $n_{\rm sd}\times n_{\rm sd}$ real skew-symmetric matrices $\mathfrak{so}(\mathbb{R},n_{\rm sd})$.


When $n_{\rm sd}=3$ we can associate with $\usf{W}(\tau)$ the matrix
$\boldsymbol{\sf w}(\tau) := \left(w_i(\tau)\right)_{i\in\mathcal{I}}\in \mathcal{M}_{3,1}(\mathbb{R})$ that is defined such that
\begin{equation}
\boldsymbol{\mathsf{W}}(\tau)\boldsymbol{\sf x}=\boldsymbol{\sf w}(\tau) \times\boldsymbol{\sf x}
\label{eq:SkewSymProp}
\end{equation}
for all $\boldsymbol{\sf x}\in \mathcal{M}_{3,1}(\mathbb{R})$. In $\eqref{eq:SkewSymProp}$, the symbol ``$\times$'' denotes the (Gibbs) ``cross-product'' so that  $\boldsymbol{\sf w}(\tau)\times\boldsymbol{\sf x} = \pr{\epsilon_{ijk} w_j(\tau) x_k}_{i\in \mathcal{I}}$. Here,  $\epsilon_{ijk}$ is  the Levi-Civita symbol. It is $\pm 1$ when $(i,j,k)$ is, respectively, an even or odd permutation of $(1,2,3)$ and is zero otherwise.
The relation between $\usf{W}(\tau)$ and $\usf{w}(\tau)$ can also be expressed using the map $\star{\left(\cdot\right)}:   \mathfrak{so}(\mathbb{R},3) \to \mathcal{M}_{3,1}(\mathbb{R})$ that is  defined by the equation $\star\pr{\cdot}=\pr{-\epsilon_{ijk}\pr{\cdot}_{jk}/2}_{i\in \mathcal{I}}$. Thus, $\usf{w}(\tau)=\star\pr{\usf{W}(\tau)}$. It can be shown that $\star^{-1}\pr{}: \mathcal{M}_{3,1}(\mathbb{R}) \to \mathfrak{so}(\mathbb{R},3)$ satisfies the equation $\star^{-1}\pr{\cdot}=\pr{-\epsilon_{ijk}\pr{\cdot}_{k}}_{i, j\in \mathcal{I}}$. To make some of the ensuing expressions appear less cumbersome we  denote  $\star^{-1}\pr{\cdot}$ too as $\star \pr{\cdot}$. Whether we mean $\star^{-1}\pr{\cdot}$ or $\star\pr{\cdot}$ will be clear from the argument of $\star\pr{\cdot}$.

Using  $\boldsymbol{\sf w}(\tau)$, $\eqref{eq:SpatialVelocityFieldMatForm}$ can alternatively be written as

\begin{equation}
\boldsymbol{\sf v}_\tau(\boldsymbol{\sf x}) = \boldsymbol{\sf w}(\tau) \times\left(\boldsymbol{\sf x}-\boldsymbol{\sf c}(\tau)\right) +\boldsymbol{\sf c}'(\tau).
\label{eq:ModSpatialVelocityFieldMatForm}
\end{equation}


\paragraph*{Accelerations}

Defining
\begin{align}
A_{\tau i}(\boldsymbol{X})=\boldsymbol{A}_{\tau}(\boldsymbol{X})\cdot_{\mathbb{A}} \boldsymbol{a}_i,\label{eq:AtauiDef}
\intertext{and}
a_{\tau i}(\boldsymbol{x}):=\boldsymbol{a}_{\tau}(\boldsymbol{x})\cdot_{\mathbb{A}} \boldsymbol{a}_i
\end{align}
it follows from $\eqref{eq:MatAccField}$, and $\eqref{eq:SpatialAccField}$ that
\begin{equation}
A_{\tau i}(\boldsymbol{X})=Q''_{ij}(\tau)X_j+c''_i(\tau),
\label{eq:MatAccFieldCoeff}
\end{equation}
and

\begin{align}
a_{\tau i}(\boldsymbol{x})&= Q''_{ik}(\tau)Q_{lk}(\tau) (x_l-c_l(\tau)) +c''_i(\tau).
\label{eq:SpatialAccFieldCoeff}
\end{align}
Denoting $\left(A_{\tau i}(\boldsymbol{X})\right)_{i\in\mathcal{I}}$, $\left(Q''_{ij}(\tau)\right)_{i,j\in\mathcal{I}}$, $\left(X_i\right)_{i\in\mathcal{I}}$, $\left(c''_i(\tau)\right)_{i\in\mathcal{I}}$, and $\left(a_{\tau i}(\boldsymbol{x})\right)_{i\in\mathcal{I}}$   as, respectively, $\boldsymbol{\mathsf{A}}_\tau(\boldsymbol{\mathsf{X}})$, $\boldsymbol{\mathsf{Q}}''(\tau)$, $\boldsymbol{\mathsf{X}}$, $\boldsymbol{\mathsf{c}}''(\tau)$, and $\boldsymbol{\sf a}_\tau(\boldsymbol{\sf x})$ and defining

\begin{equation}
\boldsymbol{\sf W}'(\tau)=\left(W'_{ij}(\tau)\right)_{i,j\in\mathcal{I}},
\label{eq:ssWPrime}
\end{equation}
where $W'_{ij}$ is the derivative of $W_{ij}$, we can write $\eqref{eq:MatAccFieldCoeff}$, and $\eqref{eq:SpatialAccFieldCoeff}$ alternatively as

\begin{equation}
\boldsymbol{\mathsf{A}}_\tau(\boldsymbol{\mathsf{X}})=\boldsymbol{\mathsf{Q}}''(\tau)\boldsymbol{\mathsf{X}}+\boldsymbol{\mathsf{c}}''(\tau),
\label{eq:MatAccFieldMatrixForm}
\end{equation}
and

\begin{align}
    \boldsymbol{\sf a}_\tau(\boldsymbol{\sf x}) = \boldsymbol{\sf W}'(\tau)\left(\boldsymbol{\sf x}-\boldsymbol{\sf c}(\tau)\right) + \boldsymbol{\sf W}^2(\tau)\left(\boldsymbol{\sf x}-\boldsymbol{\sf c}(\tau)\right) +\boldsymbol{\sf c}''(\tau),
\label{eq:SpatialAccFieldMatForm}
\end{align}
respectively. In $\eqref{eq:SpatialAccFieldMatForm}$ $\boldsymbol{\sf W}^2(\tau):= \boldsymbol{\sf W}(\tau) \boldsymbol{\sf W}(\tau)$. In arriving at  $\eqref{eq:SpatialAccFieldMatForm}$ we used the identity

\begin{align}
Q''_{ik}(\tau)Q_{lk}(\tau) = W'_{il}(\tau)+W_{ik}(\tau)W_{kl}(\tau),
\label{eq:SpatialAngAccCoeff}
\end{align}
which follows from $\eqref{eq:OrthonormalityCondition}$.

It can be shown that $\boldsymbol{\sf W}'(\tau)$ belongs to $\mathfrak{so}(\mathbb{R},n_{\rm sd})$. Therefore, when $n_{\rm sd}=3$ it follows from $\eqref{eq:SkewSymProp}$ that we can construct a matrix $\boldsymbol{\sf w}'(\tau)\in \mathcal{M}_{3,1}(\mathbb{R})$ such that $\boldsymbol{\sf W}'(\tau)\boldsymbol{\sf x}=\boldsymbol{\sf w}'(\tau)\times \boldsymbol{\sf x}$ for all $\boldsymbol{\sf x}\in \mathcal{M}_{3,1}(\mathbb{R})$. Using $\boldsymbol{\sf w}'(\tau)$, $\eqref{eq:SpatialAccFieldMatForm}$ can alternatively be written as

\begin{equation}
\boldsymbol{\sf a}_\tau(\boldsymbol{\sf x}) = \boldsymbol{\sf w}'(\tau)\times \left(\boldsymbol{\sf x}-\boldsymbol{\sf c}(\tau)\right) + \boldsymbol{\sf w}(\tau)\times\boldsymbol{\sf w}(\tau)\times \left(\boldsymbol{\sf x}-\boldsymbol{\sf c}(\tau)\right) +\boldsymbol{\sf c}''(\tau).
\label{eq:SpatialAccMatrix}
\end{equation}
We call the second term in \eqref{eq:SpatialAccMatrix}, $\boldsymbol{\sf w}(\tau)\times\boldsymbol{\sf w}(\tau)\times \left(\boldsymbol{\sf x}-\boldsymbol{\sf c}(\tau)\right)$, the ``non-linear rotational acceleration.''\footnote{The second term in \eqref{eq:SpatialAccMatrix}, $\boldsymbol{\sf w}(\tau)\times\boldsymbol{\sf w}(\tau)\times \left(\boldsymbol{\sf x}-\boldsymbol{\sf c}(\tau)\right)$,
is sometimes referred to as ``centripetal acceleration''~\cite{franck2015extracting,tan2005design,zou2014isotropic}.}

\paragraph*{Pseudo accelerations}

Defining
\begin{equation}
    \bar{A}_{\tau i}(\boldsymbol{X}):=\bar{\boldsymbol{A}}_\tau(\boldsymbol{X})\cdot_{\mathbb{E}_{\rm R}}\boldsymbol{E}_i,
\label{eq:barAtauiDef}
\end{equation}
and expressing
 $\boldsymbol{P}_\tau$, $\boldsymbol{q}(\boldsymbol{\tau})$, and $\u{X}$, respectively, as  $P_{ij}(\tau)\boldsymbol{E}_i\otimes\boldsymbol{E}_j$,  $q_i(\tau)\boldsymbol{E}_i$, and $X_i\u{E}_i$ it follows from~\eqref{eq:BarAcc} that
 \begin{align}
\bar{\boldsymbol{\sf A}}_\tau(\boldsymbol{\sf X}) =  \boldsymbol{\sf P}(\tau)\boldsymbol{\sf X} +\boldsymbol{\sf q}(\tau),
\label{eq:BarAccMatrixForm}
\intertext{where $\bar{\boldsymbol{\sf A}}_\tau(\boldsymbol{\sf X})=\left(\bar{A}_{\tau i}(\boldsymbol{X})\right)_{i\in\mathcal{I}}$, $\boldsymbol{\sf P}(\tau) =\left(P_{ij}(\tau)\right)_{i,j\in\mathcal{I}}$, and $\boldsymbol{\sf q}(\tau) = \left(q_i(\tau)\right)_{i\in\mathcal{I}}$, and from \eqref{eq:PtauMap},~\eqref{eq:Qbar}, and~\eqref{eq:MtauDef} that }
\usf{P}(\tau)= \usf{Q}^{\sf T}(\tau)\, \usf{Q}''(\tau),
\label{eq:PtauMat}
\end{align}
where $\usf{Q}''(\tau):=\left(Q''_{ij}(\tau)\right)_{i,j\in\mathcal{I}}$.




\subsection{Balance principles}
\label{sec:GeneralBalanceLawDerivation}

Consider the measurable space $(\mathcal{B},\boldsymbol{B})$, where $\boldsymbol{B}$ is the Borel $\sigma$-algebra of $\mathcal{B}$. Let $\tilde{\rho}$ be a measure on this space that is defined such that the mass contained in any $U\in \boldsymbol{B}$ is equal to $\tilde{\rho}(U)m$, where $m$ is the total mass of $\mathcal{B}$. We call $\tilde{\rho}$ the specific mass measure. A large variety of rigid bodies can be modeled through the use of $\tilde{\rho}$.  For example, in our framework the rigid solid can be an assembly of point masses, 1D or 2D straight or curved continua, such as links, rings, trusses, plates, shells, etc., and 3D solids (cf.~\cite[p. 38]{truesdell2004non}).

The position vectors of $\mathcal{B}$'s center of mass in $\mathbb{E}_{\rm R}$ and $\mathbb{E}$ are
$$
\boldsymbol{r}_{\rm R}:=\int_{\mathcal{B}}\boldsymbol{\kappa}_{\rm R}\, d\tilde{\rho},
$$
and
$$
\boldsymbol{r}(\boldsymbol{\tau}):=\int_{\mathcal{B}}\boldsymbol{\kappa}_{\tau}\, d\tilde{\rho},
$$
respectively. The vector $\boldsymbol{r}(\boldsymbol{\tau})$ can be written as $r_i(\tau) \boldsymbol{e}_i$, where $r_i:\mathbb{R}\rightarrow\mathbb{R}$.

We model the force system acting on $\mathcal{B}$ using the vector valued measure
 $\tilde{\boldsymbol{f}}_{\tau}:\boldsymbol{B}\rightarrow\mathbb{A}$ where  $\tilde{\boldsymbol{f}}_{\tau}(U)$ gives the net force acting on $U$ divided by $\mathcal{B}$'s total mass. The specific force $\tilde{\boldsymbol{f}}_{\tau}(U)$ can be expressed as $\tilde{f}_{\tau i}(U)\boldsymbol{a}_i$, where $\tilde{f}_{\tau i}$ is a signed measure on $(\mathcal{B},\boldsymbol{B})$.
 Modeling the force system as a measure allows us to model complicated loading scenarios. For example, if the concentrated force $ m \tilde{\boldsymbol{p}}  $, where  $\tilde{\boldsymbol{p}} \in \mathbb{A}$, constantly acts on the particle $\mathcal{X}$ then in that case $\tilde{\boldsymbol{f}}_{\tau}=\delta_{(\mathcal{X})} \tilde{\boldsymbol{p}}$, where $\delta_{(\mathcal{X})}: \boldsymbol{B} \to \mathbb{R}$ is called the Dirac measure and is defined as
 \begin{align}
\delta_{(\mathcal{X})}(U)=
\left\{
\begin{array}{ll}
1, & {\mathcal{X}}\in U,\\
0, & \text{otherwise}.
\end{array}
\right.
\label{eq:DiracMeasureDef}
\end{align}
Say the force on a particle of mass $M$ due to gravity is  $ M \mathscr{g} \boldsymbol{a}_3$, where $\mathscr{g}\in \mathbb{R}$, then the loading  due to gravity on $\mathcal{B}$ can be modeled as $\tilde{\boldsymbol{f}}_{\tau}= \mathscr{g} \tilde{\rho}\boldsymbol{a}_{3}$\footref{UnitsGravity}.
The specific force measure $\sum_{i=1}^{n}\delta_{(\mathcal{X}_i)}\tilde{\boldsymbol{p}}_i+
\mathscr{g} \tilde{\rho}\boldsymbol{a}_{3}$ corresponds to a scenario where the body is subject to both gravitational loading and the set of concentrated specific forces $\tilde{\boldsymbol{p}}_i\in \mathbb{A}$, where $i=1,\ldots,n$, that act, respectively, on the particles $\mathcal{X}_i$.

It follows from the principle of balance of linear momentum that
\begin{equation}
 \boldsymbol{r}''(\boldsymbol{\tau}) = \tilde{\boldsymbol{f}}_{\tau}(\mathcal{B}),
\label{eq:LMB}
\end{equation}
where $\tilde{\boldsymbol{f}}_{\tau}(\mathcal{B}) = \tilde{f}_{\tau i}(\mathcal{B}) \boldsymbol{a}_i$ is the net specific force acting on $\mathcal{B}$,  and $\boldsymbol{r}''$ is the second (Fr\'echet) derivative of the map $\mathbb{T}\ni \boldsymbol{\tau}\mapsto \boldsymbol{r}(\boldsymbol{\tau})\in \mathbb{E}$. Expressing $\boldsymbol{r}''(\boldsymbol{\tau})$ as $r''_i(\tau) \boldsymbol{a}_i$, where $r''_i$ is the second derivative of $r_i$, \eqref{eq:LMB} can be alternatively written as
\begin{equation}
     {\boldsymbol{\sf r}}''(\tau) = \tusf{f}(\tau),
\label{eq:MatrixFormLMB}
\end{equation}
where ${\boldsymbol{\sf r}}''(\tau) = \left(r''_i(\tau)\right)_{i\in\mathcal{I}}$ and $\tusf{f}(\tau):=\left(\tilde{f}_{\tau i}(\mathcal{B})\right)_{i\in\mathcal{I}}$.

Let $\boldsymbol{\Pi}_{\rm R}:\mathcal{B}\rightarrow\mathbb{E}_{\rm R}$,  and $\boldsymbol{\pi}_{\tau}:\mathcal{B}\rightarrow\mathbb{E}$ be maps defined by the equations
\begin{equation}
\boldsymbol{\Pi}_{\rm R}(\mathcal{X}) = \boldsymbol{\kappa}_{\rm R}(\mathcal{X})-\boldsymbol{r}_{\rm R},
\label{eq:PosVecRefConfigCM}
\end{equation}
and
\begin{equation}
\boldsymbol{\pi}_{\tau}(\mathcal{X}) = \boldsymbol{\kappa}_{\tau}(\mathcal{X})-\boldsymbol{r}(\boldsymbol{\tau}),
\label{eq:PosVecDefConfigCM}
\end{equation}
respectively. We call  $\boldsymbol{\Pi}_{\rm R}(\mathcal{X})$ and $\boldsymbol{\pi}_\tau(\mathcal{X})$  the relative position vectors of  $\mathcal{X}$ in the reference space and the physical space, respectively.


 It follows from the definitions of $\boldsymbol{\kappa}_{{\rm R}}$, $\boldsymbol{\kappa}_{\tau}$,  $\boldsymbol{r}_{{\rm R}}$, and $\boldsymbol{r}_{\tau}$,
and \eqref{eq:DefCompos}, \eqref{eq:PosVec},  \eqref{eq:PosVecRefConfigCM}, and
\eqref{eq:PosVecDefConfigCM} that
\begin{align}
\boldsymbol{\pi}_\tau(\mathcal{X}) = \boldsymbol{Q}_{\tau} \boldsymbol{\Pi}_{{\rm R}}(\mathcal{X}),
\label{eq:RelPosVecDefRefConfigCM}
\end{align}
 which in component form reads
\begin{align}
\pi_{\tau i}(\mathcal{X}) = Q_{ij}(\tau) \Pi_{{{\rm R}} j}(\mathcal{X}).
\label{eq:RelPosVecDefRefConfigCoeff}
\end{align}

In $\eqref{eq:RelPosVecDefRefConfigCoeff}$, $\Pi_{{{\rm R}} i}$ and $\pi_{\tau i}$ are functions from  $\mathcal{B}$ into $\mathbb{R}$ that are defined by the equations  $\Pi_{{\rm R} i}(\mathcal{X})=\boldsymbol{\Pi}_{{\rm R}}(\mathcal{X})\cdot \boldsymbol{E}_i$ and  $\pi_{\tau i}(\mathcal{X})=\boldsymbol{\pi}_{\tau}(\mathcal{X})\cdot \boldsymbol{e}_i$, respectively.

We model $\mathcal{B}$'s specific angular momentum and the net specific torque acting on it as    $\tilde{H}_{ij}(\tau)\boldsymbol{e}_i\otimes \boldsymbol{v}_j$  and
$ \tilde{T}_{ij}(\tau)\boldsymbol{e}_i\otimes\boldsymbol{a}_j$, respectively. Here $\tilde{H}_{ij}$, and $\tilde{T}_{ij}$ are real valued functions on $\mathbb{R}$ defined by the equations
\begin{equation}
\tilde{H}_{ij}(\tau)=  \int_{\mathcal{B}} \Big(\dot{\pi}_{\tau i}\pi_{\tau j} - \pi_{\tau i} \dot{\pi}_{\tau j}\Big) \, d\tilde{\rho},
\label{eq:AngMomentumTensorCoeff}
\end{equation}
and
\begin{equation}
\tilde{T}_{ij}(\tau)  = \int_{\mathcal{B}}
\left(\pi_{\tau j} \, d\tilde{f}_{\tau i} - \pi_{\tau i} \, d\tilde{f}_{\tau j}\right),
\label{eq:TorqueTensorCoeff}
\end{equation}
respectively. In \eqref{eq:AngMomentumTensorCoeff}, $\dot{\pi}_{\tau i}:\mathcal{B}\to \mathbb{R}$ is defined by the equation
\begin{equation}
\dot{\pi}_{\tau i}(\mathcal{X})=\pi'_{\mathcal{X} i}(\tau),
\label{eq:DotPiSubTau}
\end{equation}
where $\pi_{\mathcal{X} i}'$ is the derivative of $\pi_{\mathcal{X}i}$,  and $\pi_{\mathcal{X} k}:\mathbb{R}\to \mathbb{R}$ is   defined by the equation
\begin{equation}
\pi_{\mathcal{X} k}(\tau)=\pi_{\tau k}(\mathcal{X}).
\label{eq:PisubChi}
\end{equation}

It follows from the principle of balance of angular momentum that
\begin{equation}
\tusf{H}'(\tau) = \tusf{T}(\tau),
\label{eq:DefAMBtensorCoeff}
\end{equation}
where $\tusf{H}(\tau):=(\tilde{H}_{ij}(\tau))_{i, j\in \mathcal{I}}$, $\tusf{H}'(\tau):=(\tilde{H}'_{ij}(\tau))_{i, j\in \mathcal{I}}$, in which $\tilde{H}'_{ij}$ is the derivative of $\tilde{H}_{ij}$, and $\tusf{T}(\tau) := (\tilde{T}_{ij}(\boldsymbol{\tau}))_{i,j\in\mathcal{I}}$.

In \S
\ref{sec:DerDefAMBmatrixFormPrecursor} we show that
\begin{equation}
    \tusf{H}'(\tau)=\usf{W}'(\tau) \tusf{E}(\tau)+ \tusf{E}(\tau) \usf{W}'(\tau) +  \usf{W}^2(\tau) \tusf{E}(\tau) - \tusf{E}(\tau) \usf{W}^2(\tau),
\label{eq:DefAMBmatrixFormPrecursor}
\end{equation}
where $ \boldsymbol{\sf W}$ and $\boldsymbol{\sf W}'$ are defined via  \eqref{eq:SpatialAngularVelocityTensorCoeff}, \eqref{eq:ssW}, and \eqref{eq:ssWPrime}, and $\tilde{\boldsymbol{\sf E}}(\tau) := (\tilde{E}_{kj}(\tau))_{k,j\in\mathcal{I}}$, in which
\begin{equation}
\tilde{E}_{ij}(\tau) := \int_{\mathcal{B}} \pi_{\tau i}\pi_{\tau j} \, d\tilde{\rho}.
\label{eq:EulerTensorDefConfigCoeff}
\end{equation}


\subsubsection{The case \texorpdfstring{$n_{\rm sd}=3$}{nsd=3}}

The matrices appearing in the principle of balance of angular momentum \eqref{eq:DefAMBtensorCoeff} belong to $\mathfrak{so}(\mathbb{R}, n_{\rm sd})$. Hence, \eqref{eq:DefAMBtensorCoeff}  represents only $n_{\rm sd}$ number of independent equations. This fact can be made  more explicit when $n_{\rm sd}=3$ by writing ~\eqref{eq:DefAMBtensorCoeff} in the following alternate form.

When $n_{\rm sd}=3$, we can define
\begin{align}
\tilde{\boldsymbol{\mathsf{h}}} (\tau)&=  \star\pr{\tilde{\boldsymbol{\mathsf{H}}} (\tau)},
\label{eq:hiDef}
\intertext{and}
\tilde{\usf{t}} (\tau)&=\star \pr{\tilde{\boldsymbol{\mathsf{T}}} (\tau)}.
\label{eq:tiDef}
\end{align}
In \S\ref{sec:nsd3details} we show that it follows from  ~\eqref{eq:DefAMBtensorCoeff},~\eqref{eq:DefAMBmatrixFormPrecursor},~\eqref{eq:hiDef}, and~\eqref{eq:tiDef}  that
\begin{equation}
\tilde{\boldsymbol{\sf J}}(\tau) \boldsymbol{\sf w'}(\tau) + \boldsymbol{\sf W}(\tau)  \tilde{\boldsymbol{\sf J}}(\tau)\boldsymbol{\sf w}(\tau)  = \tilde{\boldsymbol{\sf t}}(\tau),
\label{eq:AMBcolumnMat}
\end{equation}
where $\usf{w}(\tau)=\star\pr{ \usf{W}(\tau)}$, $\tilde{\boldsymbol{\sf J}}(\tau) := \left(\tilde{J}_{ij}(\tau)\right)_{i,j\in\mathcal{I}}$, in which
\begin{equation}
\tilde{J}_{ij}(\tau) := \int_{\mathcal{B}} \Big(\pi_{\tau k} \pi_{\tau k}\delta_{i j} - \pi_{\tau i} \pi_{\tau j}\Big) \, d\tilde{\rho}.
\label{eq:InertiaTensorsDefConfigCoeff}
\end{equation}

\paragraph*{Modified angular momentum balance }

The balance of angular momentum ~\eqref{eq:AMBcolumnMat} is a non-linear differential equation that can be solved analytically only for simple cases. In \S
\ref{sec:RigidBodySimulations} we discuss a procedure for solving~\eqref{eq:AMBcolumnMat} numerically. We use the numerical solutions to~\eqref{eq:AMBcolumnMat} for generating the virtual accelerometer data that we use for testing our algorithm described in \S\ref{Sec:PA}. The numerical procedure we employ is based on the following alternate formulation of the balance of angular momentum statement because the dependence of the inertia matrix on time,  $\tilde{J}(\tau)$, in~\eqref{eq:AMBcolumnMat} makes effecting a numerical solution  to~\eqref{eq:AMBcolumnMat}  quite challenging.

Using~\eqref{eq:RelPosVecDefRefConfigCoeff} in \eqref{eq:InertiaTensorsDefConfigCoeff} we can show that
\begin{equation}
\tilde{\usf{J}}(\tau)
=\usf{Q}(\tau)\,\barspJMat\,\usf{Q}^{\sf T}(\tau),
\label{eq:RelInertiaTensorsRefDefConfig}
\end{equation}
where $\barspJMat:=\left(\bar{\tilde{J}}_{i j}\right)_{i,\,j\in\mathcal{I}}$, with $\bar{\tilde{J}}_{i j}$ being defined by the equation
\begin{equation}
\bar{\tilde{J}}_{i j} = \int_{\mathcal{B}} \Big(\Pi_{{\rm R} k} \Pi_{{\rm R} k} \delta_{i j} - \Pi_{{\rm R} i}  \Pi_{{\rm R} j}\Big) \, d\tilde{\rho}
~(\text{no sum over ${\rm R}$}).
\label{eq:InertiaTensorsRefConfigCoeff}
\end{equation}
Substituting $\tilde{\boldsymbol{\sf J}}(\tau)$ with the right hand side of \eqref{eq:RelInertiaTensorsRefDefConfig} in \eqref{eq:AMBcolumnMat}, and then multiplying both sides of the resulting equation with $\boldsymbol{\sf Q}^{\sf T}(\tau)$ we get
\begin{equation}
\barspJMat\,\bar{\usf{w}}'(\tau) + \bar{\usf{ W}}(\tau)\,\barspJMat\,\bar{\usf{ w}}(\tau)  = \usf{Q}^{\sf T}(\tau) \,\tilde{\usf{t}}(\tau),
\label{eq:ModAMBcolumnMat}
\end{equation}
where
\begin{equation}
    \bar{\usf{W}}(\tau):=\tpsb{\usf{Q}}(\tau)\, \usf{W}(\tau)\,\usf{Q}(\tau),
\label{eq:barWMatDef}
\end{equation}
$\bar{\boldsymbol{\sf w}}(\tau):=\star \pr{\bar{\usf{W}}(\tau)}$ and $\bar{\usf{w}}'$ is $\bar{\usf{w}}$'s derivative.

Since $\barspJMat$ does not depend on time, ~\eqref{eq:ModAMBcolumnMat} is relatively much easier to  solve than~\eqref{eq:AMBcolumnMat}. Once $\bar{\usf{w}}$ is known from the solution of~\eqref{eq:ModAMBcolumnMat},  $\usf{w}$ can be computed from
\begin{equation}
\boldsymbol{\sf w}(\tau) =\boldsymbol{\sf Q}(\tau)\,\bar{\boldsymbol{\sf w}}(\tau),
\label{eq:wFROMwBar}
\end{equation}
which is a consequence of~\eqref{eq:barWMatDef} and the definition of $\bar{\boldsymbol{\sf w}}(\tau)$.

\section{Proposed algorithm}
\label{Sec:PA}

\begin{figure}[ht!]
  \centering
\includegraphics[width=\textwidth]{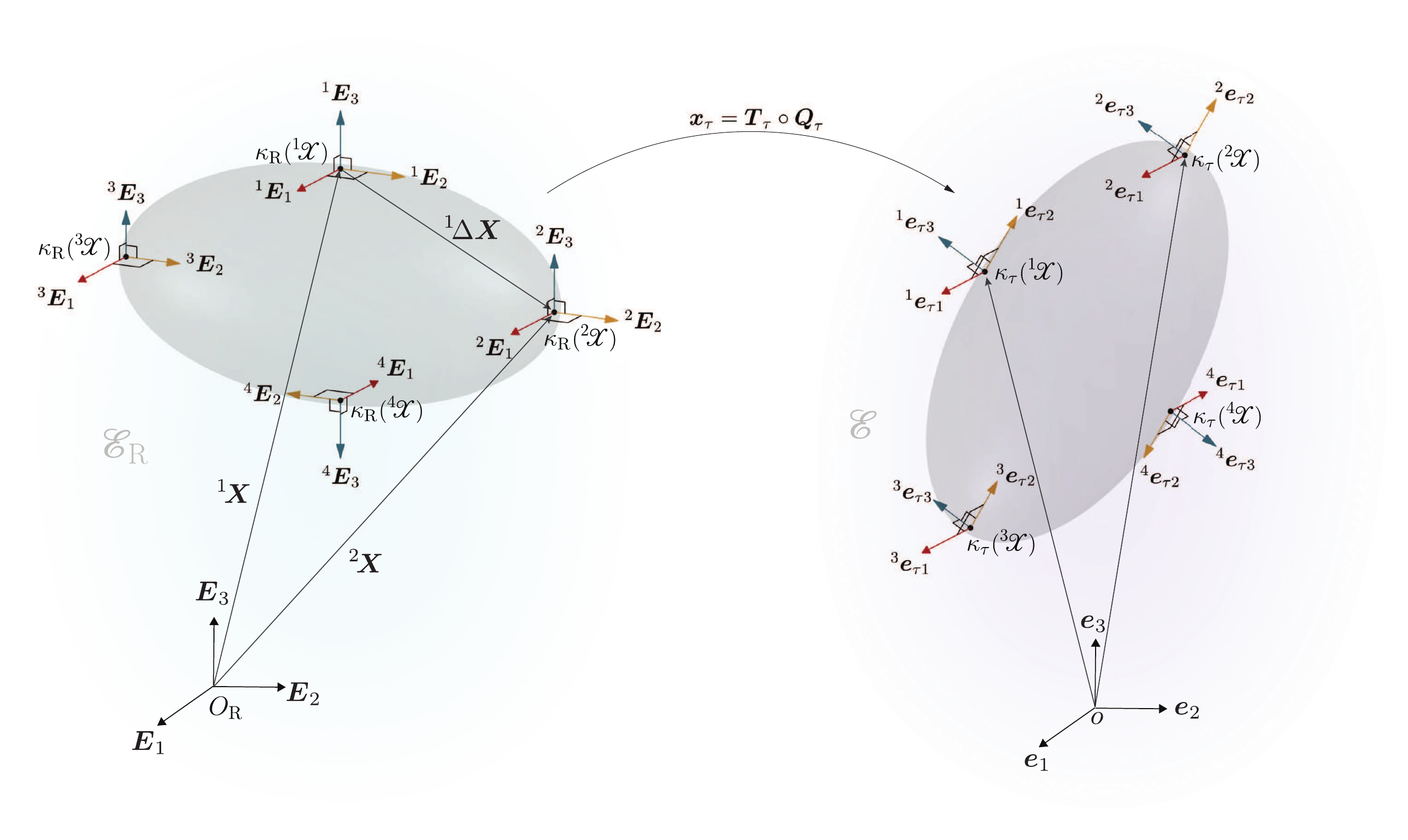}
\caption{Location and orientation of accelerometers. The set of vectors with which the accelerometer $\lsc{\u{X}}$'s material axes are parallel at the time instance $\tau$ are given by $\left(\lsc{\u{e}}_{\tau i}\right)_{i\in \mathcal{I}}$.
}
\label{fig:AccelerometerFigure}
\end{figure}

We propose an algorithm for determining the acceleration field $\boldsymbol{A}_{\rm \tau}:\boldsymbol{\kappa}_{\rm R}(\mathcal{B})\to \mathbb{A}$ using measurements from four tri-axial accelerometers. We consider the accelerometers themselves to be rigid bodies and to be attached to $\mathcal{B}$. We number the accelerometers $1,\ldots, 4$, respectively, and denote the material particles that they are attached to as  $\lsc{\mathcal{X}}$ or $\lsc{\u{X}}$, where, of course, $\lsc{\u{X}}=\boldsymbol{\kappa}_{\rm R}(\lsc{\mathcal{X}})$ and $\mathscr{l}\in \mathcal{J}:=(1, \ldots, 4)$ denotes an accelerometer's number (see Fig.~\ref{fig:AccelerometerFigure}).

A tri-axial  accelerometer is capable of measuring the components of its acceleration (or equivalently the acceleration of the material particle that it is attached to) in three  orthonormal directions. These directions, in our formalism, belong to $\mathbb{A}$  and depend on the three mutually perpendicular material  axes of the accelerometer along which it is said to report its acceleration's components. The accelerometers are rigidly attached to $\mathcal{B}$ and move with it (see Fig.~\ref{fig:AccelerometerFigure}). By which we mean that as $\mathcal{B}$ executes its motion in $\mathbb{E}$,  the accelerometers attached to it also move  with it and change their orientations. We denote the set of vectors in $\mathbb{E}$ with which the accelerometer $\lsc{\u{X}}$'s material axes are parallel at the time instance $\tau$ as $\left(\lsc{\u{e}}_{\tau i}\right)_{i\in \mathcal{I}}$.  Owing to the accelerometer $\lsc{\u{X}}$'s rigid nature, it follows that $\left(\lsc{\u{e}}_{\tau i}\right)_{i\in \mathcal{I}}$ is orthonormal, i.e. $\lsc{\u{e}}_{\tau i}\cdot\lsc{\u{e}}_{\tau j}=\delta_{ij}$ (no sum over left superscript $\mathscr{l}$), and from the rigid nature of $\mathcal{B}$ and $\lsc{\u{X}}$'s  attachment to it, that the vector set  $\left(\lsc{\u{E}}_{ i}\right)_{i\in \mathcal{I}}$ where

\begin{equation}
\lsc{\u{E}}_i:=
\boldsymbol{Q}_{\tau}^*\,\,\, \lsc{\u{e}}_{\tau i}
\label{eq:mathscrlEiDef}
\end{equation}
is also orthonormal and, more importantly, constant with time.

\subsection{Procedure for computing the pseudo acceleration}
\label{Sec:CompPseudoAccAtMatPoint}

The accelerometer  $\lsc{\u{X}}$ reports the functions  $\lsc{\alpha}_i:\mathbb{R}\to \mathbb{R}$ such that its acceleration

\begin{equation}
\u{A}_\tau(\lsc{\u{X}})=\lsc{\alpha}_{ i}(\tau)
\lsc{\u{a}}_{\tau i}~\text{(no sum over left superscript $\mathscr{l}$)}.
\label{eq:AccAtAccelemeterPosition}
\end{equation}
Here, the acceleration vectors $\lsc{\u{a}}_{\tau i}\in \mathbb{A}$ are defined by the physical vectors $\lsc{\u{e}}_{\tau i}$. To be precise, $\lsc{\u{a}}_{\tau i}$ are defined such that $\left(\lsc{\u{a}}_{\tau i}\u{s}\right)\u{s}= \lsc{\u{e}}_{\tau i}$, where recall that $\u{s}$ is a fixed vector in $\mathbb{T}$ of unit norm. Substituting $\lsc{\u{a}}_{\tau i}$ with $Q_{jk}(\tau) \left(\u{E}_k\cdot\lsc{\u{E}}_{i}\right)\boldsymbol{a}_j$ in   \eqref{eq:AccAtAccelemeterPosition},  operating both sides of the resulting equation with $\bar{\u{Q}}_\tau$, which is defined in~\eqref{eq:Qbar}, and identifying    the expression $\bar{\boldsymbol{Q}}_\tau\circ\boldsymbol{A}_\tau$ that appears on the  left hand side as $\bar{\boldsymbol{A}}_\tau$ we get that

\begin{equation}
\bar{\u{A}}_\tau(\lsc{\u{X}}) = \lsc{\alpha}_{ i}(\tau)\lsc{\u{E}}_{i} \,(\textrm{no sum over left superscript}\, \mathscr{l}).
\label{eq:FicAccAtAcclerometerPoistion}
\end{equation}
It follows from \eqref{eq:BarAccMatrixForm}, and \eqref{eq:FicAccAtAcclerometerPoistion} that

\begin{equation}
\lsc{\bar{\usf{A}}}(\tau) = \boldsymbol{\sf P}(\tau) \,\lsc{\usf{X}} + \boldsymbol{\sf q}(\tau),
\label{eq:BarAccAccPosMatrixForm}
\end{equation}
where  $\lsc{\bar{\usf{A}}}(\tau):=
\left(\bar{\u{A}}_{\tau}(\lsc{\u{X}})\cdot
\u{E}_i\right)_{i\in\mathcal{I}}$, and $\lsc{\usf{X}} :=\left({}^{\mathscr{l}}\!\boldsymbol{X}\cdot \u{E}_i\right)_{i\in\mathcal{I}}$. It can be shown using  \eqref{eq:BarAccAccPosMatrixForm} that

\begin{equation}
\lsc[i]{\Delta\bar{\usf{A}}}(\tau)=\usf{P}(\tau) \, \lsc[i]{\Delta \usf{X}},
\label{eq:DiffBarAccAccPosMatrixForm}
\end{equation}
where $\lsc[i]{\Delta\bar{\usf{A}}}(\tau):=\lsc[i+1]{\bar{\usf{A}}}(\tau)-\lsc[1]{\bar{\usf{A}}}(\tau)$, $\lsc[i]{\Delta\usf{X}}:=\lsc[i+1]{\usf{X}}-\lsc[1]{\usf{X}}$, and $i\in \mathcal{I}$.


Solving \eqref{eq:DiffBarAccAccPosMatrixForm} we get

\begin{equation}
\usf{P}(\tau)
= \pr{
\tps{\lsc[i]{\Delta\bar{\usf{A}}}(\tau)}\, \pr{\lsc[j]{\dsf{X}}}}\,\pr{\pr{\lsc[i]{\dsf{Y}}}
\,
\tps{\lsc[j]{\dsf{Y}}}
},
\,
\label{eq:PmatSol}
\end{equation}
where $\lsc[i]{\Delta\usf{Y}}$ are defined such that

\begin{equation}
\tps{\lsc[i]{\Delta\usf{Y}}}\lsc[j]{\Delta\usf{X}}=\delta_{ij},\quad i,~j\in \mathcal{I}.
\end{equation}

For the case $n_{\rm sd}=3$ it can be shown that
\begin{equation}
\lsc[i]{\Delta\usf{Y}}=\frac{1}{2} \epsilon_{ijk}
\left(
\frac{\pr{\lsc[j]{\Delta\usf{ X}}}\times\pr{\lsc[k]{\Delta\usf{ X}}}}{\tps{\lsc[1]{\Delta\usf{ X}}}
\left(\pr{\lsc[2]{\Delta\usf{X}}}\times \pr{\lsc[3]{\Delta\usf{X}}}\right)
}
\right).
\label{eq:InverseBasisMat}
\end{equation}
Knowing $\boldsymbol{\sf P}$ from~\eqref{eq:PmatSol}, the function $\usf{q}$ can be obtained from~\eqref{eq:BarAccAccPosMatrixForm} as
\begin{equation}
\usf{q}(\tau)=
\lsc{\bar{\usf{A}}}(\tau) - \boldsymbol{\sf P}(\tau) \,\lsc{\usf{ X}},
\label{eq:qmatSol}
\end{equation}
where $\mathscr{l}$ can be any particular integer in $\mathcal{J}$. Knowing $\boldsymbol{\sf P}$ and $\boldsymbol{\sf q}$ from \eqref{eq:PmatSol} and \eqref{eq:qmatSol}, respectively,  $\bar{\boldsymbol{\sf A}}_\tau(\usf{X})$ is now completely determined by
\eqref{eq:BarAccMatrixForm}.

\subsection{Procedure for computing the acceleration from pseudo acceleration}
\label{sec:ProcPseudoAcc}

It follows from \eqref{eq:barAtauDef}, \eqref{eq:Qbar}, \eqref{eq:AtauiDef}, \eqref{eq:barAtauiDef}, and \eqref{eq:OrthonormalityConditionMatrixForm} that

\begin{equation}
\boldsymbol{\sf A}_\tau(\boldsymbol{\sf X})= \boldsymbol{\sf Q}(\tau)\bar{\boldsymbol{\sf A}}_\tau(\boldsymbol{\sf X}).
\label{eq:RelAccAndAccBar}
\end{equation}
Once $\usf{Q}$ is known  the acceleration $\usf{A}_{\tau}(\usf{X})$ can be computed from $\bar{\usf{A}}_{\tau}(\usf{X})$ using~\eqref{eq:RelAccAndAccBar}. Therefore, in the following, we discuss the procedure for computing $\usf{Q}$.


Differentiating \eqref{eq:OrthonormalityConditionMatrixForm} twice, adding $\tpsb{\usf{Q}}(\tau)\usf{Q}''(\tau)$ on both sides of the resulting equation, and rearranging we get that

\begin{equation}
\usf{Q}^{\sf T}(\tau)\,\usf{Q}''(\tau) + \left(\usf{Q}^{\sf T}\right)'(\tau)\,\usf{Q}'(\tau) = \frac{1}{2}\left(\usf{Q}^{\sf T}(\tau)\,\usf{Q}''(\tau)-\left(\usf{Q}^{\sf T}\right)''(\tau)\,\usf{Q}(\tau)\right).
\label{eq:DDofOrthiMod}
\end{equation}
Using \eqref{eq:DDofOrthiMod} and \eqref{eq:PtauMat}, it can be shown that

\begin{equation}
\bar{\usf{W}}'(\tau)=\textrm{skew}\left(\usf{P}(\tau)\right),
\label{eq:SkewP}
\end{equation}
where $\textrm{skew}(\cdot):= \pr{\pr{\cdot}-\tps{\cdot}}/2$. Integrating both sides of \eqref{eq:SkewP} we get

\begin{equation}
\bar{\boldsymbol{\sf W}}(\tau)=\int_{z=0}^{\tau}\textrm{skew}\left(\boldsymbol{\sf P}(z)\right)\,dz + \bar{\boldsymbol{\sf W}}(0).
\label{eq:IntSkewP}
\end{equation}
Knowing $\bar{\usf{W}}'$ and $\bar{\usf{W}}$  from~\eqref{eq:SkewP} and~\eqref{eq:IntSkewP}, respectively, $\usf{Q}$ can be obtained by solving the differential equation
\begin{equation}
\usf{Q}'(\tau)=\usf{Q}(\tau)\bar{\usf{W}},
\label{eq:EulerEquation}
\end{equation}
which is a consequence of  the definitions of $\usf{W}$ and $\bar{\usf{W}}$.  We propose the following numerical procedure for solving~\eqref{eq:EulerEquation}.

We denote the value of  $\usf{Q}$ at the  time instance $n \Delta \tau$ as  $\usf{Q}(n)$. In this definition the symbol $\Delta \tau$ denotes the discrete (non-dimensional) time increment and  is a positive real number,  and $n$ is the time step and is a non-negative  integer. Applying the  Lie-St\"{o}rmer-Verlet numerical integration scheme~\cite{terze2015angular} to~\eqref{eq:EulerEquation}, knowing $\usf{Q}(n)$, the value of $\usf{Q}$ at the next  time step  can be computed as

\begin{equation}
\boldsymbol{\sf Q}(n+1) = \boldsymbol{\sf Q}(n)\, e^{\Delta\tau\,\bar{\boldsymbol{\sf W}}_{n+1/2}},
\label{eq:UpdateRotMat}
\end{equation}
where

\begin{equation}
\bar{\usf{W}}_{n+1/2} := \bar{\usf{W}}(n)+ \frac{\Delta \tau}{2} \bar{\usf{W}}'(n),
\label{eq:WbarMatNthPlusHalfStep}
\end{equation}
and the map $\mathsf{e}^{\pr{\cdot}}:\mathfrak{so}(\mathbb{R},3)\to SO(3)$ is defined by the equation
\begin{equation}
\mathsf{e}^{\pr{\cdot}} = \usf{I}+ \text{sinc}\left(\frac{\left\lVert \star \pr{\cdot}\right\rVert}{2}\right)\pr{\cdot} + \frac{1}{2}\left[\text{sinc}\left(\frac{\left\lVert {\star \pr{\cdot}}\right\rVert}{4}\right)\right]^2 \pr{\cdot}^2.
\label{eq:ExpSkewSymm}
\end{equation}
In~\eqref{eq:WbarMatNthPlusHalfStep} the symbols $\busf{W}'(n)$ and $\busf{W}(n)$, respectively, denote the values of  $\busf{W}'$ and $\busf{W}$ at the time instance $n \Delta\tau$, and $\lVert\cdot \rVert$ returns the $\ell^2$-norm.
These values can be computed using~\eqref{eq:SkewP} and~\eqref{eq:IntSkewP}, respectively.

\section{Validation of the \emph{accelerometer only} (AO) algorithm using virtual accelerometer data}
\label{Sec:NumericaResults}

\begin{figure}
\centering
\includegraphics[width=0.9\textwidth]{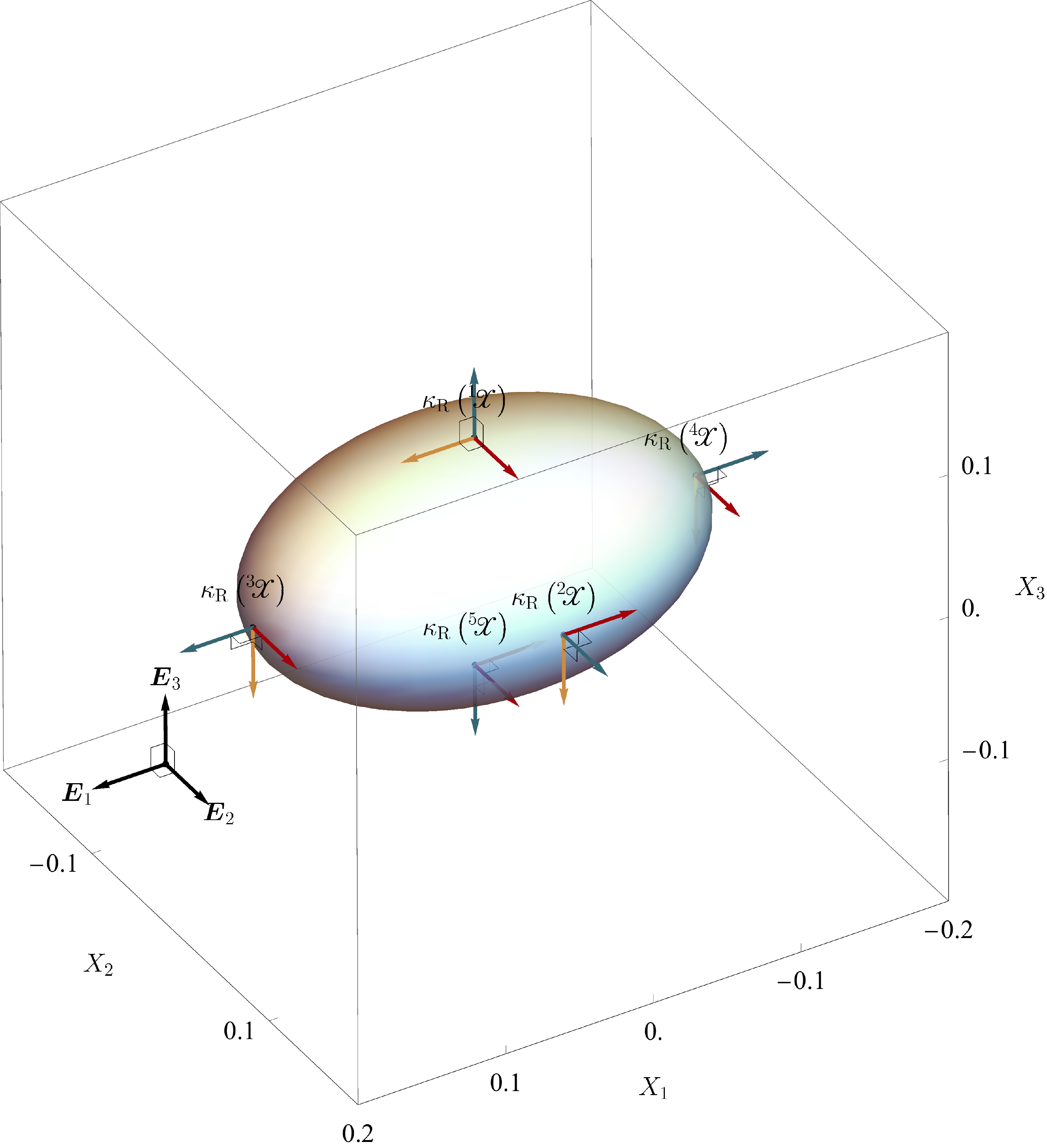}
\caption{Accelerometer arrangement in numerical simulations of impact between a rigid body and an elastic half-space (see \S\ref{Sec:NumericaResults} for details). In the numerical simulations we take $\mathcal{B}$ to be a rigid ellipsoid.
    In the reference point space $\mathcal{E}_{\rm R}$  the ellipsoid occupies the region  $\{(X_1,X_2,X_3)\in \mathcal{E}_{\rm R}~|~\left(X_1/a\right)^2+\left(X_2/b\right)^2+\left(X_3/c\right)^2=1\}$, where $a=0.15$, $b=0.10$, and $c=0.08$. Four accelerometers are attached to the ellipsoid's material  particles $\lsc{\mathcal{X}}$, where $\mathscr{l}\in \mathcal{J}$. The reference position vectors of these particles are, respectively, $(h + c)\boldsymbol{E}_3$, $b\boldsymbol{E}_2+h\boldsymbol{E}_3$, $a\boldsymbol{E}_1+h \boldsymbol{E}_3$, and $h\boldsymbol{E}_3-a\boldsymbol{E}_1$, where $h=0.75$.
    We apply the AO-algorithm to the accelerometer data supplied by the virtual accelerometers $\lsc{\mathcal{X}}$, $\mathscr{l}\in \mathcal{J}$, to predict the acceleration of the material particle $\lsc[5]{\mathcal{X}}$.
		We  add varying amounts of noise to the virtual accelerometer data before applying the AO-algorithm to it.
		The material particle $\lsc[5]{\mathcal{X}}$'s  reference position vector is  $(h-c)\boldsymbol{E}_3$.}
\label{fig:EllipsoidAccPos}
\end{figure}

To apply the AO-algorithm we need acceleration measurements from four tri-axial accelerometers.
As a preliminary step we checked the validity and robustness of the AO-algorithm using virtual accelerometer data. We generated these data by performing a numerical simulation of rigid body motion, extracting the virtual accelerometer measurements from that simulation,  and adding varying amounts of noise to those measurements. 
In the future we plan to validate the AO-algorithm using experimentally generated accelerometer data.

\subsection{Generation of virtual accelerometer data}
\label{sec:VAD}

\subsubsection{Numerical simulation of rigid body motion}
\label{sec:NSimRBM}

We generated the virtual accelerometer data by numerically simulating impact between a rigid body and an elastic half-space. In the simulation we took  $\boldsymbol{E}_i$, $i\in \mathcal{I}$, to have units of meters and $\boldsymbol{s}$ to have units of seconds. This implied that $\boldsymbol{v}_i$ and $\boldsymbol{a}_i$, $i\in \mathcal{I}$, had units of meters-per-second and meters-per-second-squared, respectively.
In the simulation we took  $\mathcal{B}$ to be a rigid ellipsoid.
The ellipsoid was spatially and directionally homogeneous, by which we mean that for all $U\in \boldsymbol{B}$,  where, recall, $\boldsymbol{B}$ is $\mathcal{B}$'s Borel $\sigma$-algebra, $\tilde{\rho}(U)$ was  equal to the ratio between the Lebesgue measures of $\kappa_{\rm R}(U)$ and $\kappa_{\rm R}(\mathcal{B})$.
The ellipsoid was fitted with the  number and type of accelerometers needed for the application of the AO-algorithm. The configuration of $\mathcal{B}$  and the arrangement of its accelerometers in $\mathcal{E}_{\rm R}$  is shown in Fig.~\ref{fig:EllipsoidAccPos}. The position and orientation of the surface of the elastic half-space $H$ when there was no contact between it and $\mathcal{B}$ is  shown in Fig.~\ref{fig:VirtualAccData} (a).
In the simulation the ellipsoid was prescribed initial angular and translational velocities and dropped onto the  elastic half-space under the action of gravity. 
Specifically, the initial conditions in the simulation were $\busf{w}_0=(5,5,5)$,  $\usf{r}'_0 = (0.75,0,0)$,  $\usf{r}_0 = (0,0,0.75)$, and $\usf{Q}_0=\text{the identity element in $\mathcal{M}_{3,3}(\mathbb{R})$}$.
The Young's modulus and Poisson's ratio of the elastic half space $H$ in the simulation were $10^4$Pa and $0.3$, respectively.
The  value of the ellipsoid's specific force measure on a $U\in \boldsymbol{B}$ consisted of two parts. The first part was constant with time and approximately equal to $-9.8\tilde{\rho}(U) \boldsymbol{a}_3$\footnote{
\label{UnitsGravity}
Note that the units of this value are carried by $\boldsymbol{a}_3$}.
The second part arose from $\mathcal{B}$'s interaction with  $H$. We discuss the calculation of this second part in \S
\ref{sec:CompForTorduImp}.

A few representative configurations of the ellipsoid from the simulation are shown in Fig.~\ref{fig:VirtualAccData}(a).
We discuss the details of the numerical scheme that we used for performing the simulation in  \S
\ref{sec:RigidBodySimulations}.
Further details about the simulation can be found in Fig.~\ref{fig:VirtualAccData}.
The  data from    accelerometer $\lsc[1]\mathcal{X}$ in the simulation are shown in Fig.~\ref{fig:VirtualAccData} (b).
We detail the  procedure that we used for extracting the data from the virtual accelerometers  in the simulation in \S
\ref{sec:VirtualAccelerometerData}.

\subsubsection{Adding noise to the virtual accelerometer data}
\label{sec:NoiseAddition}

To make the data from the virtual accelerometers more representative of experimental accelerometer data, we added different amounts of noise to it before feeding it to the AO-algorithm. 
Typically, the noise present in accelerometer measurements is modeled as band-limited additive white Gaussian noise (AWGN) \cite{park2006error,liu2014design}. 
The Ornstein-Uhlenbeck (OU) process is a good model for band-limited AWGN~\cite[ch. 2]{dimian2014noise}~\cite{bibbona2008ornstein}. 
Therefore, we  generated noisy virtual accelerometer data $\lsc{\alpha}_i^{\rm Noisy}$ from the virtual accelerometer data ${}^{\mathscr{l}}\!\alpha_i$ from the simulation as
$$
\lsc{\alpha}_i^{\rm Noisy}(\tau) = \lsc{\alpha}_i(\tau) + \eta_{\tau},
\label{eq:NoisyAccData}
$$
where $\eta_{\tau}$ is a particular realization of the  OU process. 
The OU process is a continuous time and  state stochastic process that satisfies the  integral equation
\begin{equation}
\eta_{\tau_1+\tau_2}-\eta_{\tau_1}
=-\beta \int_{\tau_1}^{\tau_1+\tau_2}\eta_{\tau}\,
  d\tau + \sigma\int_{\tau_1}^{\tau_1+\tau_2} \, dW_{\tau},
  \label{eq:OUprocess}
\end{equation}
in which the second integral is an It\^{o} integral,  $W_{\tau}$ is the Wiener process,  $\beta>0$ is the drift coefficient, $\sigma \geq 0$ is the diffusion coefficient, and $\tau_1,~\tau_2$ are time instances. 
A particular realization of the OU process is obtained  by drawing $\eta_0$ from a Gaussian distribution of mean zero  and variance $\sigma^2/(2\beta)$ and solving~\eqref{eq:OUprocess}. 
A part of the noisy virtual accelerometer data that we generated for the case $\beta=10^3$ and $\sigma=100$ is shown in Fig.~\ref{fig:VirtualAccData} (c). 

The  values of the power spectral density (PSD) of the noise in  commercial accelerometers appear to remain approximately constant over the  100 Hz frequency band~\cite{rogers1997accelerometer}. 
The order of magnitude of the mean of those values is typically $10^{-8}  \rm g^2/Hz$, where g is acceleration due to gravity~\cite{ADXL345,ADXL325,rogers1997accelerometer}.
The noise is typically higher in the out of plane direction than the in-plane directions~\cite{ADXL345}. 
We found the values of the PSD of the OU process to remain approximately constant over the 100 Hz frequency band and be greater than  $10^{-8}  \rm~g^2/Hz$ and $10^{-6} \rm~g^2/Hz$ when $\sigma\ge 1$ and $\sigma\ge 10$, respectively,  and $\beta=10^3$ (see Fig.~\ref{fig:PSDplotForDiffSigmaBeta}). 
Therefore, we studied the  effect of noise on the predictive capability of the AO-algorithm for $\sigma=1,~10,$ and $100$ while keeping $\beta$ constant at $10^3$. 

\begin{figure}[ht]
   \centering
   \setlength{\unitlength}{0.1\textwidth}
   \begin{picture}(10,13.5)
     \put(0,0){\includegraphics[width=\textwidth]{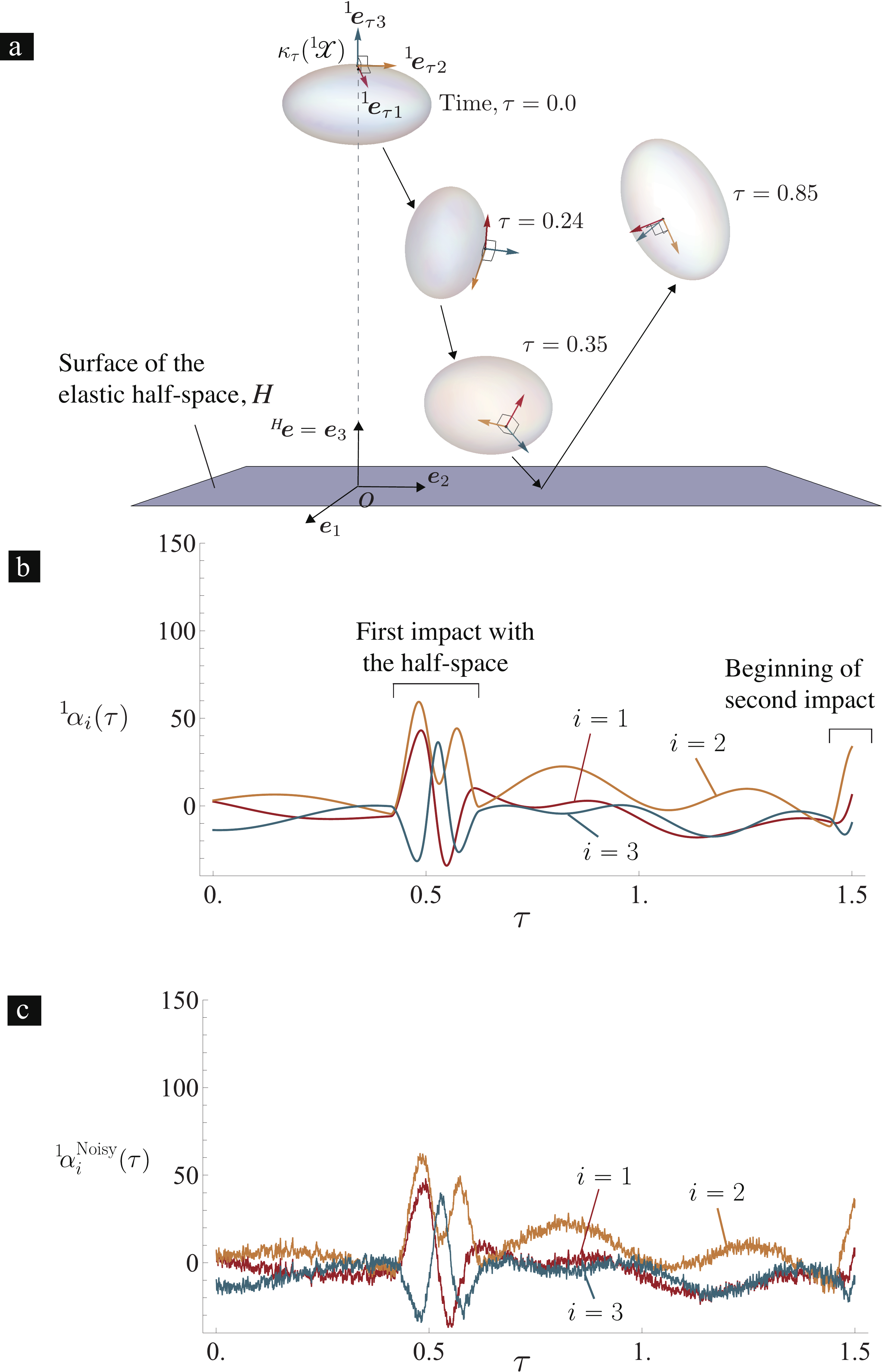}}
     \put(4.9,3.65){
     $\lsc[1]{\alpha}_i^{\rm Noisy}(\tau) = \lsc[1]{\alpha}_i(\tau) + \eta_{\tau}$}
	\put(5.0,3.2){
    $\eta_{\tau}$=~\parbox[t]{4cm}{
	\begin{flushleft}
    {\fontfamily{ptm}\selectfont  \normalsize
	Ornstein-Uhlenbeck (OU) process~\eqref{eq:OUprocess} for $\beta=10^3$ and $\sigma=100$}
	\end{flushleft}}
	}
	\end{picture}
	\caption{(Caption continued on next page)}
	   \label{fig:VirtualAccData}
\end{figure} 

\clearpage
\begin{figure}
    \centering
  \contcaption{Simulation of impact between a rigid ellipsoid and an elastic half-space. 
	The simulation is introduced in \S
    \ref{sec:NSimRBM}. 
	(a) Configurations of the ellipsoid at different time instances in the simulation. 
	The reference configuration of the ellipsoid is shown in Fig.~\ref{fig:EllipsoidAccPos}. 
	In the simulation the ellipsoid was prescribed initial angular and translational velocities and dropped onto an elastic half-space under the action of gravity. 
	Specifically, the initial conditions in the simulation were $\busf{w}_0=(5,5,5)$,  $\usf{r}'_0 = (0.75,0,0)$,  $\usf{r}_0 = (0,0,0.75)$, and $\usf{Q}_0=\text{the identity element in $\mathcal{M}_{3,3}(\mathbb{R})$}$. 
	Further details of the simulation can be found  in \S
    \ref{sec:RigidBodySimulations}. 
 (b) The components of the acceleration of the virtual accelerometer  ${}^1\!\materialPtManifold$ (see subfigure (a)) with respect to the directions ${}^1\boldsymbol{e}_{\tau i} = \boldsymbol{Q}_{\tau}\lsc[1]{\boldsymbol{E}}_i$, $i\in \mathcal{I}$, in the simulation. 
	We discuss the details of the  procedure that we used for extracting these components from the  simulation results in  \S
    \ref{sec:VirtualAccelerometerData}. 
	(c) Noisy, virtual accelerometer data generated by adding a particular  realization of the Ornstein-Uhlenbeck process~\eqref{eq:OUprocess} to the virtual accelerometer data shown in subfigure (b). 
	}
\end{figure}

\subsection{Evaluation of AO-algorithm's capability in predicting the accelerations in the simulation}

We  model noise as the OU process, which is a stochastic process. 
Therefore, for a given noise level, i.e., fixed $\sigma$ and $\beta$ values, there is no unique noisy virtual accelerometer data set. 
Instead, there exists an entire family of data sets. 
In the parlance of stochastic processes, each of those data sets is called a realization. 
Using the virtual accelerometer data from the simulation,
for each noise level we created a large number of realizations (see Table.~\ref{tb:Errors}). 
We applied the AO-algorithm individually to each of those realizations to derive separate predictions for the acceleration of the material particle $\lsc[5]{\mathcal{X}}$ in the simulation. 
The location of  $\lsc[5]{\mathcal{X}}$ in the ellipsoid's reference configuration is shown in Fig.~\ref{fig:EllipsoidAccPos}. 
We know the exact values of $\lsc[5]{\mathcal{X}}$'s acceleration from the simulation results. 
Recall that $\lsc[5]{\mathcal{X}}$'s acceleration $\boldsymbol{A}_{\tau}\pr{\lsc[5]{\boldsymbol{X}}}$ is
related to the matrix $\lsc[5]{\usf{A}}(\tau)\in \mathcal{M}_{3,1}(\mathbb{R})$  as $\pr{\lsc[5]{\usf{A}(\tau)}}_i\boldsymbol{a}_i=\boldsymbol{A}_{\tau}\pr{\lsc[5]{\boldsymbol{X}}}$.  
Figures~\ref{fig:PredAccEllipsoidTriad} (a)--(b), respectively, show the acceleration components $\pr{\lsc[5]{\usf{A}}(\tau)}_{i}$, $i\in \mathcal{I}$, as functions of time. 
Each subfigure shows both the exact values as well as the predicted values that  resulted from the application of the AO-algorithm to single realizations of noisy, virtual accelerometer data set for the noise levels $\sigma=1,~10$, and 100. 
In each of those noise levels $\beta$ was equal to $10^3$.  

As can be noted from Fig.~\ref{fig:PredAccEllipsoidTriad}, in the absence of noise, denoted by the noise level $\sigma=0$, the predicted and exact values are indistinguishable. 
Even in the presence of noise the differences between the predicted and exact values become visible  only at the noise level $\sigma=100$. 
Recall, that $\sigma$ for the noise in commercial accelerometers is typical around unity and rarely exceeds 10. 

In the simulation, the ellipsoid impacts the half-space repeatedly.
The ellipsoid first impacts the half-space and bounces back from it during the  $(0.25, 0.75)$ time interval.  
The second impact takes place around $\tau =1.5$ (see Fig.~\ref{fig:VirtualAccData} (b)). 
We found the time variations of the acceleration components for each impact to be qualitatively similar. 
However, we found the difference between the predicted and exact values to increase with time. 
This is expected due to the presence of the integration steps in the AO-algorithm (see~\eqref{eq:IntSkewP}, \eqref{eq:UpdateRotMat}, and \eqref{eq:WbarMatNthPlusHalfStep}). 
In order to make a more quantitative comparison between the predicted and exact values we decided to focus on the $[0,1]$ time frame and use the following error measures:
\begin{align}
\epsilon_p&=\frac{\lVert \text{AO}\left(\lsc[5]\usf{A}\right)-\lsc[5]\usf{A}\rVert_p} {\lVert \lsc[5]\usf{A}\rVert_p},
\label{eq:RelL2Error}
\end{align}
where $p=2$ and $\infty$.  
In~\eqref{eq:RelL2Error} the norms $\lVert \cdot\rVert_{2}$ and $\lVert \cdot\rVert_{\infty}$ are defined as $\lVert f\rVert_{2}=\sqrt{\int_{0}^{1}f(\tau) \, d\tau}$ and $\lVert f\rVert_{\infty}=\sup_{\tau\in [0,1]}\left|f(\tau)\right|$. 
The function $\text{AO}\pr{\lsc[5]{\usf{A}}}$ is defined such that $\text{AO}\pr{\lsc[5]{\usf{A}}}(\tau)$ equals the value predicted by the AO-algorithm for $\lsc[5]{\usf{A}}(\tau)$. 
It follows from~\eqref{eq:RelL2Error} that $\epsilon_{p}$ depends on both the noise level as well as on the particular realization of the noisy, virtual accelerometer data set at that noise level that was used for generating $\text{AO}\pr{\lsc[5]{\usf{A}}}$. 
Hence, we computed $\epsilon_{p}$ for a large number of realizations at each noise level and calculated  their mean and standard deviation. 
We report the results of these calculations in Table.~\ref{tb:Errors} 

As be noted from Table.~\ref{tb:Errors}, for the type of noise that is typically seen in commercial accelerometers
the error measure $\epsilon_2$ (resp. $\epsilon_{\infty}$) is  rarely going to exceed $1.5\%$ (resp. $2\%$). 
It is interesting to note that even for the case of $\sigma=100$, the error measures  $\epsilon_2$ and $\epsilon_{\infty}$ will, respectively, rarely exceed $14\%$ and  $19\%$. 
However, it might not be advisable to read too much into the values of  $\epsilon_2$ and $\epsilon_{\infty}$ since they  depend strongly on the  details of our evaluation methodology. 
To elaborate, our choice to focus on the time interval $[0,1]$ in defining $\epsilon_p$ was arbitrary. 
So was our choice for $\lsc[5]{\mathcal{X}}$'s location, or for that matter most details in the simulation. 
The values of the error measures would likely be different if we had made different choices. 
Similarly, the error measures' values also critically depend on the manner in which we  modeled the noise in the accelerometer measurements. 
For example, it is not unreasonable to imagine that the dominant noise in the accelerometer measurements does not arise from the electronic noise that is intrinsic to the accelerometers, which is what we decided to presently focus on, but from factors such as insufficiently rigid mounting of the accelerometers,  non-negligible deformations, etc. 
For these reasons, the error measures $\epsilon_{p}$ should be  used only as preliminary gauges into the robustness of the AO-algorithm.  

The  values summarized in Table.~\ref{tb:Errors} for the case $\sigma=0$  prove that the AO-algorithm provides a valid strategy to estimate the accelerations at arbitrary material particles of a rigid body using measurements from only four tri-axial accelerometers.
Despite its validity, it is reasonable to question: how practical is the AO-algorithm? 
Will it be  robust in the field in the presence of noise? 
 The values summarized in Table.~\ref{tb:Errors} tempt us to answer these questions in the affirmative. 
 However, owing to the various limitations of the error measures $\epsilon_p$ a true answer to these questions will have to wait for the results of an experimental evaluation.


\begin{figure}
\centering
\includegraphics[width=\textwidth]{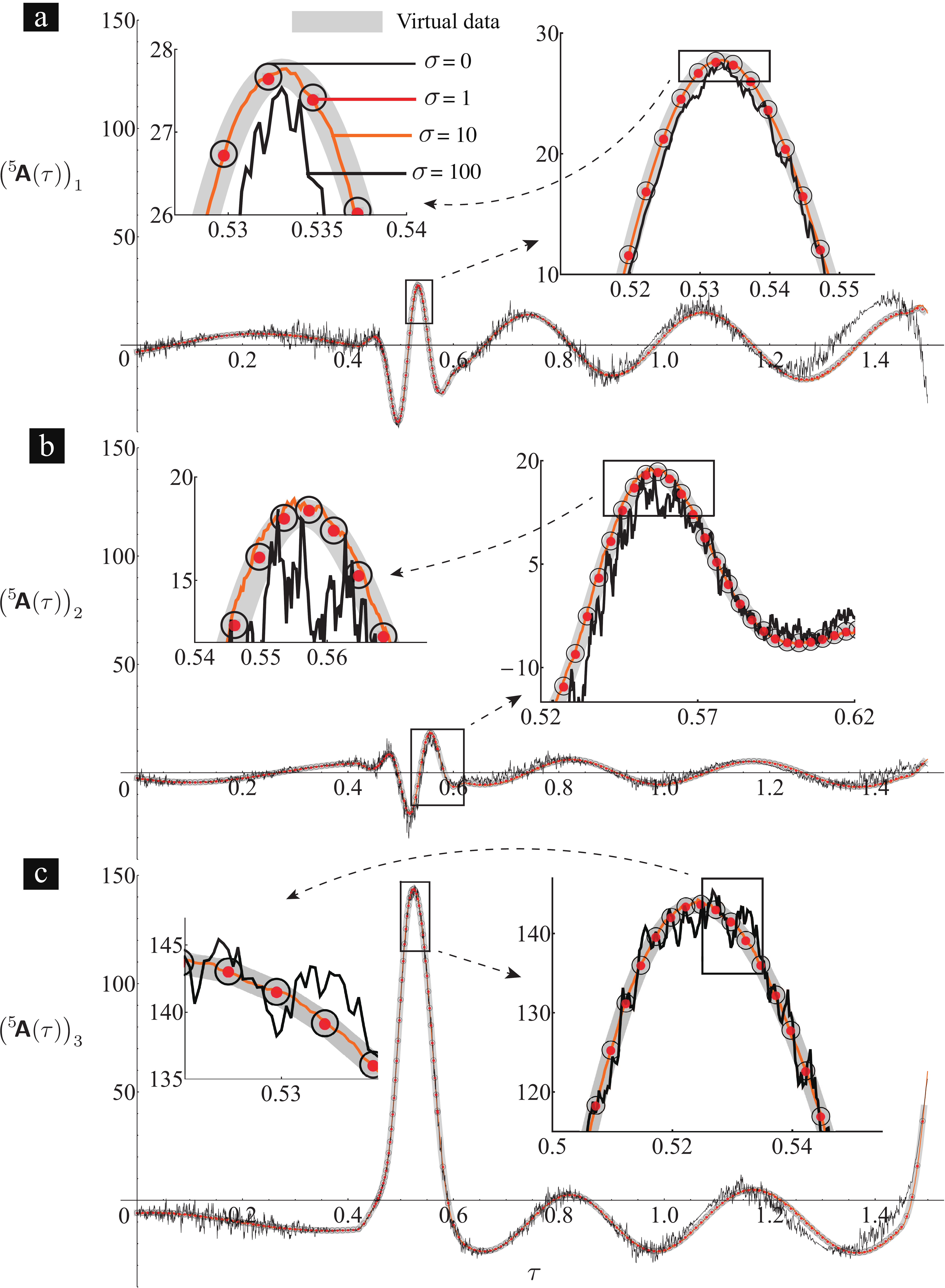}
\caption{
(Caption continued on next page) 
}
\label{fig:PredAccEllipsoidTriad}
\end{figure}


\begin{figure}[t]
  \contcaption{
  The AO-algorithm's prediction for the acceleration of the material particle $\lsc[5]{\c{X}}$ in the simulation of a rigid ellipsoid impacting an elastic half-space. 
  The location of $\lsc[5]{\c{X}}$ on the ellipsoid's surface is shown in Fig.~\ref{fig:EllipsoidAccPos}. 
  The simulation is introduced in \S\ref{sec:NSimRBM}, and discussed further in~\ref{sec:NumericalSimulation}. 
Representative results from the simulation are shown in Fig.~\ref{fig:VirtualAccData} (a). 
  Noise is addded to the accelerations of four virtual accelerometers that are taken to be attached to the ellipsoid during the simulation and fed into the AO-algorithm to derive the predictions.  
The arrangement of the four virtual accelerometers is shown in Fig.~\ref{fig:EllipsoidAccPos}. 
  The components of the acceleration time signal from the virtual accelerometer $\lsc[1]{\c{X}}$ before and after addition of a noise realization are shown in  Fig.~\ref{fig:VirtualAccData} (b) and (c), respectively. 
  The details of extraction of the virtual accelerometers' data is discussed in \S\ref{sec:VirtualAccelerometerData}. 
  Noise in our numerical validation exercise is dictated by the parameters $\sigma$ and $\beta$. 
  See \S\ref{sec:NoiseAddition} for an explanation of these parameters. 
  The figure shows the predictions for the cases $\sigma=0,~1,~10$, and $100$. 
  The parameter $\beta$ was kept fixed at $10^3$ in all these cases.}
\end{figure}




\begin {table}[H]
  \caption {The mean and standard deviation of the   error measures $\epsilon_2$ and $\epsilon_{\infty}$ for 100 and 200 realizations.}
\begin{center}
\scalebox{1.0}{%
\begin{tabular}{|c|c|c|c|c|}
\hline
\multirow{3}{*}{$\sigma$}
& \multicolumn{2}{c|}{Realizations (sample size) = 100 }
& \multicolumn{2}{c|}{Realizations (sample size) = 200 }\\
\cline{2-5}
 & $\epsilon_2 \times 10^3$
 & $\epsilon_\infty \times 10^3$
 & $\epsilon_2 \times 10^3$
 & $\epsilon_\infty \times 10^3$   \\
 & (mean$\pm$std)
 & (mean$\pm$std)
 & (mean$\pm$std)
 & (mean$\pm$std) \\
\hline
 0  & 1.56
 & 2.20
 & 1.56
 & 2.20\\
\hline
 1 & 1.83 $\pm$ 0.20
 & 2.88$\pm$ 0.31
 & 1.80$\pm$ 0.19
 & 2.85$\pm$ 0.28\\
 \hline
 10 & 8.37$\pm$ 1.60
 & 12.80$\pm$ 1.82
 & 8.42$\pm$ 1.94
 & 12.88$\pm$ 2.51 \\
 \hline
 100 & 83.99$\pm$ 18.84
 & 127.44$\pm$ 23.78
 & 82.31$\pm$ 16.68
 & 126.49$\pm$ 20.33 \\
\hline
\end{tabular}}
\end{center}
\label{tb:Errors}
\end{table}

\section{Concluding remarks}
\label{sec:Discussions}

\begin{enumerate}

\item
We checked the validity of the AO-algorithm and its robustness in the presence of noise using virtual accelerometer data generated by performing a numerical simulation. In our study we chose the parameters $\sigma$ and $\beta$ to approximate noise from commercial accelerometers. We have not studied the effects of biases or noise  resulting from poor mounting of the accelerometers, etc. on the predictive capability of the AO-algorithm. We plan to study such effects in the near future.

 \item  Without employing any time integration, the AO-algorithm is still able to determine the  pseudo acceleration field $ \bar{\boldsymbol{A}}_\tau$, from which the magnitude of acceleration for all material particles is known (see \eqref{eq:NormEqual}). With time integration the AO-algorithm further yields the  direction of acceleration for all material particles.
 \item The AO-algorithm is fairly general.
\begin{enumerate}
\item  The vector sets  $\left(\lsc{\u{E}}_i\right)_{i\in \mathcal{I}}$, where $\mathscr{l}\in \mathcal{J}$, need not form a right handed  system.
 They only need to be orthonormal.
 For example, in our simulation (see Fig.~\ref{fig:EllipsoidAccPos}) leaving  $\left(\lsc[4]{\u{E}}_i\right)_{i\in \mathcal{I}}$ none of the other vector sets form a right handed system.
 \item The spatial arrangement of the accelerometers  $\lsc{\mathcal{X}}$, $\mathscr{l}\in \mathcal{J}$, can also be fairly general except for the restrictions that all four accelerometers may not lie on the same plane, no three accelerometers shall lie on the same line, and, of course, no two accelerometers shall be placed at the same point.  These restrictions follow from  the requirement that  the matrices $\lsc[i]{\Delta\usf{Y}}$  in the AO-algorithm be well defined (cf.~\eqref{eq:InverseBasisMat}).
 \item The AO-algorithm can be applied to quite a wide variety of  rigid body motions.
 The motions can involve finite rotations and be a consequence of  a complicated set of  time varying forces and torques. For example, the motion could involve \textit{point-forces}, i.e., forces that are highly localized  in space and  arise during impact.
 The primary restriction in this direction is that the sampling frequency of the accelerometers be high enough that they  are able to capture all relevant features of the acceleration time signals $\lsc{\alpha}_i$, where $i\in \mathcal{I}$ and $\mathscr{l}\in \mathcal{J}$.
\end{enumerate}

\item  The results of Padgaonkar {\it{et. al.}}\cite{padgaonkar1975measurement} can be reproduced by making a special choice for the  accelerometers' positions and orientations in the AO-algorithm.

\item Using gyroscopes and accelerometers the minimum number of measurements needed to reproduce the acceleration field is six. Our work shows that on using accelerometers only the minimum number of such measurements is twelve (cf. \eqref{eq:BarAccAccPosMatrixForm}). The AO-algorithm can be modified to use measurements from either six bi-axial or twelve uni-axial accelerometers, instead of  measurements from four tri-axial accelerometers. We plan on publishing the details of these modifications elsewhere.


 \end{enumerate}

\clearpage





\section*{Funding Information}
The authors gratefully acknowledge support from the Panther Program and the Office of Naval Research (Dr. Timothy Bentley) under grant N000141812494.
\section*{Declaration of Competing Interests}
The authors declare that each individually and collectively have no conflict of interest.
\section*{Acknowledgments}
The authors thank Weilin Deng, and Sayaka Kochiyama  for their help in preparing some of the figures in the manuscript.





\bibliography{RBD}
\bibliographystyle{ieeetr}

\newpage
\appendix


\section{Balance Laws}
\label{sec:AppenGeneralBalanceLawDerivation}
\subsection{Derivation of~\texorpdfstring{\eqref{eq:DefAMBmatrixFormPrecursor}}{(49)}}
\label{sec:DerDefAMBmatrixFormPrecursor}

Defining $\pi_{\mathcal{X} i}(\tau) = \pi_{\tau i}(\mathcal{X})$ and $\Pi_{\mathcal{X} i} =\Pi_{{\rm R} i}(\mathcal{X})$, we can alternatively write~\eqref{eq:RelPosVecDefRefConfigCoeff} as
\begin{align}
\pi_{\mathcal{X} i}(\tau) = Q_{ij}(\tau) \Pi_{\mathcal{X} j}.
\label{eq:ModRelPosVecDefRefConfigCoeff}
\end{align}
It follows from \eqref{eq:ModRelPosVecDefRefConfigCoeff} and~\eqref{eq:OrthonormalityConditiona} that
\begin{align}
\Pi_{\mathcal{X} j} = Q_{k j}(\tau) \pi_{\mathcal{X} k}(\tau).
\label{eq:RelPosVecRefDefConfigCoeff}
\end{align}
Differentiating both sides of \eqref{eq:ModRelPosVecDefRefConfigCoeff} we get the equation   $\pi'_{\mathcal{X} i}(\tau)=Q'_{ij}(\tau) \Pi_{\mathcal{X} j}$. On replacing $\Pi_{\mathcal{X} j}$ in it with the right hand side of~\eqref{eq:RelPosVecRefDefConfigCoeff}, and then noting  in the resulting equation using~\eqref{eq:SpatialAngularVelocityTensorCoeff} that $Q'_{ij}( \tau) Q_{kj}( \tau)$  is  equal to $W_{ik}( \tau)$ and replacing
$\pi_{\mathcal{X} k}(\tau)$  with $\pi_{\tau k}(\mathcal{X})$ we get that
\begin{align}
\dot{\pi}_{\tau i}(\mathcal{X}) = W_{ik}(\tau) \pi_{\tau k}(\mathcal{X}),
\label{eq:RatePosVecDefConfigCoeff}
\end{align}
where $\dot{\pi}_{\tau i}(\mathcal{X}):=\pi'_{\mathcal{X} i}(\tau)$.

It can be shown using \eqref{eq:EulerTensorDefConfigCoeff} and~\eqref{eq:RatePosVecDefConfigCoeff} that
\begin{equation}
\tilde{E}'_{ij}(\tau) = W_{ik}(\tau)\tilde{E}_{kj}(\tau) - \tilde{E}_{ik}(\tau) W_{kj}(\tau).
\label{eq:ModRateEulerTensorDefConfigCoeff}
\end{equation}
It follows from \eqref{eq:AngMomentumTensorCoeff}, \eqref{eq:EulerTensorDefConfigCoeff}, and~\eqref{eq:RatePosVecDefConfigCoeff} that
\begin{equation}
\tilde{H}_{ij}(\tau)= W_{ik}(\tau) \tilde{E}_{kj}(\tau) +  \tilde{E}_{ik}(\tau)W_{kj}(\tau).
\label{eq:AngMomentumTensorCoeffEulerTesorCoeff}
\end{equation}
Differentiating both sides of  \eqref{eq:AngMomentumTensorCoeffEulerTesorCoeff} and then replacing the derivatives of $\tilde{E}_{ik}(\tau)$ and  $\tilde{E}_{kj}(\tau)$ on the right hand side of the resulting equation  using \eqref{eq:ModRateEulerTensorDefConfigCoeff} we get that
\begin{align}
\tilde{H}'_{ij}(\tau) &= W'_{ik}(\tau) \tilde{E}_{kj}(\tau) +  \tilde{E}_{ik}(\tau)W'_{kj}(\tau)\notag \\ &+ W_{ik}(\tau) W_{kl}(\tau) \tilde{E}_{lj}(\tau) -  \tilde{E}_{il}(\tau)W_{lk}(\tau) W_{kj}(\tau).
\label{eq:RateAngMomentumTensorCoeffEulerTesorCoeff}
\end{align}
Equation~\eqref{eq:DefAMBmatrixFormPrecursor} is the matrix form of~\eqref{eq:RateAngMomentumTensorCoeffEulerTesorCoeff}.

\subsection{Derivation of~\texorpdfstring{\eqref{eq:AMBcolumnMat}}{(53)}}
\label{sec:nsd3details}

It follows from \eqref{eq:DefAMBtensorCoeff}, \eqref{eq:hiDef}, and \eqref{eq:tiDef} that
\begin{equation}
\tilde{h}'_k(\tau) = \tilde{t}_k(\tau),
\label{eq:AMBcomponentMat}
\end{equation}
where $\tilde{h}_k(\tau):=(\tusf{h}(\tau))_k$, and $\tilde{t}_k(\tau):=(\tusf{t}(\tau))_k$.
It can be shown using \eqref{eq:InertiaTensorsDefConfigCoeff} and  ~\eqref{eq:RatePosVecDefConfigCoeff} that
\begin{align}
\tilde{J}'_{ij}(\tau)
& = W_{ik}(\tau) \tilde{J}_{kj}(\tau) - \tilde{J}_{ik}(\tau) W_{kj}(\tau).
\label{eq:RateInertiaTensorsDefConfigCoeff}
\end{align}
If follows from~\eqref{eq:AngMomentumTensorCoeff} and~\eqref{eq:hiDef} that
\begin{align}
\tilde{h}_k(\tau) &= \int_{\mathcal{B}} \epsilon_{klm}\pi_{\tau l} \dot{\pi}_{\tau m} \, d\tilde{\rho}.
\label{eq:AMcolumnMat}
\end{align}
It follows from \eqref{eq:AMcolumnMat}, \eqref{eq:RatePosVecDefConfigCoeff},~\eqref{eq:SkewSymProp},  and~\eqref{eq:InertiaTensorsDefConfigCoeff} that
\begin{align}
\tilde{h}_k(\tau)
&= \tilde{J}_{kp}(\tau) w_p(\tau).
\label{eq:AMcolumnMatCompo}
\end{align}
Differentiating~\eqref{eq:AMcolumnMatCompo} and
 using~\eqref{eq:RateInertiaTensorsDefConfigCoeff} it can be shown that
\begin{equation}
h_k'(\tau)=\tilde{J}_{km}(\tau) w'_m(\tau)+
W_{km}(\tau)\tilde{J}_{mn}(\tau) w_n(\tau).\\
\label{eq:AMBComp}
\end{equation}
In arriving at~\eqref{eq:AMBComp} we set  $W_{mp}(\tau)w_{p}(\tau)=0$, which is a consequence of~\eqref{eq:SkewSymProp}. Equation~\eqref{eq:AMBcolumnMat} follows from \eqref{eq:AMBcomponentMat} and \eqref{eq:AMBComp}.


\section{Numerical validation of the AO algorithm}
\label{sec:AppenRigidBodySimulations}

\subsection{Numerical simulation of rigid body motion}
\label{sec:NumericalSimulation}

As a preliminary step we checked the validity and robustness of the AO-algorithm using numerically generated virtual accelerometer data. We generated the virtual accelerometer data by performing simulations of rigid body motion. We discuss the details of the numerical scheme for performing the simulations in this section.

\subsubsection{Numerical integration of rigid body equations of  motion}
\label{sec:RigidBodySimulations}

We performed simulations of rigid body motion by solving the balance of linear momentum ~\eqref{eq:MatrixFormLMB} and balance of angular momentum ~\eqref{eq:ModAMBcolumnMat} simultaneously and numerically to obtain a sequence of configurations for $\mathcal{B}$. We assume that $\mathcal{B}$ updates its configuration discontinuously at a set of discrete time instances.



As in \S
\ref{sec:ProcPseudoAcc}, we assumed that the values of $\usf{c}$, $\usf{W}$, $\usf{r}$, $\usf{r}'$, $\usf{Q}$, $\busf{W}$, $\tusf{f}$, and $\tusf{t}$ remained constant during the time interval $\Delta \tau_n:=[n\Delta \tau, (n+1)\Delta \tau)$, where $n\in \mathcal{N}:=(0,1,\ldots)$, and denote them as $\usf{c}(n)$, $\usf{W}(n)$, $\usf{r}(n)$, $\usf{r}'(n)$, $\usf{Q}(n)$, $\busf{W}(n)$, $\tusf{f}(n)$, and $\tusf{t}(n)$, respectively. Recall that $\Delta \tau$ denotes the  (non-dimensional) time increment and  is a positive real number,  and $n$ is the time step (number). Knowing $\usf{r}(n)$, $\usf{r}'(n)$, $\usf{Q}(n)$, and $\busf{W}(n)$ we used the following procedure for computing  $\usf{r}(n+1)$, $\usf{r}'(n+1)$, $\usf{Q}(n+1)$, and  $\busf{W}(n+1)$.

Knowing $\usf{r}(n)$ and $\usf{Q}(n)$ and using~\eqref{eq:PosVecMatForm} we calculated $\usf{c}(n)$ as
\begin{equation}
\usf{c}(n)=\usf{r}(n)-\usf{Q}(n) \usf{r}_{\rm R}.
\label{eq:cn}
\end{equation}
Knowing $\usf{Q}(n)$ and $\usf{c}(n)$, the equation~\eqref{eq:PosVecMatForm} completely determined   $\mathcal{B}$'s configuration  for the $n^{\rm th}$ time step, or equivalently, the time interval $\Delta \tau_n$. We denote that configuration as $\boldsymbol{\kappa}_{n}:=\boldsymbol{\kappa}_{\tau},~\tau\in \Delta \tau_n$.

Since in our simulation the force measure only depends on $\mathcal{B}$'s configuration it follows that it too remains contant during each time step.
We denote that force measure as $\tu{f}_{n}:=\tu{f}_{\tau},~\tau\in \Delta \tau_n$.
We discuss the calculation of  $\tu{f}_n$ in \S\ref{sec:CompForTorduImp}.
We computed the matrices  $\tusf{f}(n)$ and $\tusf{t}(n)$ by substituting $\tu{f}_{\tau}$ with $\tu{f}_{n}$ in the definitions of $\tusf{f}$ and $\tusf{t}$, respectively.


Knowing $\tusf{f}(n)$ and applying the ``one-step formulation'' (as presented in~\cite[p. 472]{hairer2006geometric}) of the Str\"{o}mer-Verlet numerical integration scheme to~\eqref{eq:LMB} we computed $\usf{r}(n+1)$ as
\begin{equation}
{\usf{r}}(n+1) = {\usf{r}}(n) + \Delta \tau  \,\, {\usf{r}}'_{n+1/2},
\end{equation}
where
\begin{equation}
{\usf{r}}'_{n+1/2} := {\usf{r}}'(n) + \frac{\Delta \tau}{2} {\tusf{f}}(n).
\end{equation}
The formulation of the Str\"{o}mer-Verlet scheme we employed is sometimes also referred to as the ``velocity Verlet'' algorithm~\cite[p. 100]{allen2017computer}.

Knowing $\tusf{t}(n)$ and applying the explicit Lie-St\"{o}rmer-Verlet integration scheme~\cite{terze2015angular} to~\eqref{eq:EulerEquation} and~\eqref{eq:ModAMBcolumnMat}  we computed $\usf{Q}(n+1)$ as
\begin{equation}
\usf{Q}(n+1) = \usf{Q}(n)\, \usf{e}^{\Delta \tau\,\star \pr{\bar{\usf{w}}_{n+1/2}}},
\label{eq:UpdateRotMatSim}
\end{equation}
where
\begin{equation}
\bar{\usf{w}}_{n+\frac{1}{2}} = \btusf{J}^{-1}\usf{e}^{-\Delta\tau\,\bar{\usf{W}}(n)/2}\left(\btusf{J}\,\pr{\star\busf{W}(n)} + \frac{\Delta \tau}{2} \usf{Q}^{\sf T}(n)\,\tilde{\usf{t}}(n) \right).
\label{eq:AngVelColMatNthPlusHalfStep}
\end{equation}
The function  $\usf{e}^{\pr{\cdot}}$ appearing in~\eqref{eq:UpdateRotMatSim} and~\eqref{eq:AngVelColMatNthPlusHalfStep} is defined in~\eqref{eq:ExpSkewSymm}.

Knowing $\usf{r}(n+1)$ and $\usf{Q}(n+1)$, the configuration of $\mathcal{B}$  is then completely determined for the $(n+1)^{\rm th}$ time step.
We computed $\tu{f}_{n+1}$ from  $\u{\kappa}_{n+1}$ using the same procedure  that we used for computing $\tu{f}_{n}$ from  $\u{\kappa}_{n}$ (see \S\ref{sec:CompForTorduImp} for details), and  then computed $\tusf{f}(n+1)$ and $\tusf{t}(n+1)$ by substituting $\tu{f}_{\tau}$ with $\tu{f}_{n+1}$ in the definitions of $\tusf{f}$ and $\tusf{t}$, respectively.
Using those quantities, and applying the velocity-Verlet algorithm to~\eqref{eq:MatrixFormLMB}  we computed $\usf{r}'(n+1)$ as
\begin{equation}
{\usf{r}}'(n+1) ={\usf{r}}'_{n+1/2} + \frac{\Delta \tau}{2}  {\tusf{f}}(n+1).
\end{equation}

Applying the Lie-St\"{o}rmer-Verlet integration scheme  to~\eqref{eq:EulerEquation} and~\eqref{eq:ModAMBcolumnMat} we computed $\busf{W}(n+1)$ as $\star \busf{w}(n+1)$, where

\begin{equation}
\begin{aligned}
\bar{\usf{w}}(n+1) & =\btusf{J}^{-1} \usf{e}^{-\Delta\tau\,\star \pr{\busf{w}_{n+1/2}}} \ \left(\btusf{J}\,\star\pr{\busf{W}(n)}  + \frac{\Delta \tau}{2} \usf{Q}^{\sf T}(n)\,\tilde{\usf{t}}(n) \right)\\
&+  \frac{\Delta \tau}{2} \btusf{J}^{-1}\,\usf{Q}^{\sf T}(n+1)\,\tilde{\usf{t}}(n+1).
\label{eq:UpdateAngVelColMat}
\end{aligned}
\end{equation}
Following~\eqref{eq:barWMatDef} and~\eqref{eq:OrthonormalityConditionMatrixFormb},  we computed the value of $\usf{W}$ at, say, the $(n+1)^{\rm th}$ time step as $\usf{Q}(n+1)\busf{W}(n+1)\tps{\usf{Q}(n+1)}$.

\subsubsection{The force measure in a simulation time step}
\label{sec:CompForTorduImp}

Our simulation of rigid body motion involved contact between $\mathcal{B}$ and an elastic half-space. In that simulation we took $\mathcal{B}$ to be a  rigid ellipsoid, and the elastic half-space to occupy the region $H= \{o+\u{y}\in\c{E}:\u{y}\in \mathbb{E}~\text{and}~(\u{y}-\lsc[{H}]{\u{y}})\cdot\lsc[{H}]{\u{e}}\leq 0\}$. In    ${H}$'s definition the  vector $\lsc[{H}]{\u{y}}\in \mathbb{E}$ locates some arbitrary point on  $H$'s surface and the  vector $\lsc[H]{\u{e}}\in \mathbb{E}$ is the normal to $H$'s surface that points away from $H$ and is of unit norm. The  $H$ and $\lsc[{H}]{\u{e}}$ that we used in the simulation are shown in~Fig.~\ref{fig:VirtualAccData} (a).

In the simulation,  the ellipsoid's specific force measure  consisted of two parts. The first part was due to the action of gravity on $\c{B}$. Therefore,  $\tu{f}_n$ included the term $-9.8\tilde{\rho} \boldsymbol{a}_3$\footnote{
Note that the units of this value are carried by $\boldsymbol{a}_3$} for all $n\in \mathcal{N}$.
The second part arose from $\kappa_{n}\left(\mathcal{B}\right)$'s interaction with  $H$. We discuss its calculation in the remainder of this section.

For the time step $n$ we designated the material particle that is closest to the half-space as $\lsc[{H}]{\c{X}}_n=\boldsymbol{\kappa}_{n }^{-1}\pr{\lsc[H]{\u{x}}_n}$, where
\begin{equation}
\lsc[H]{\u{x}}_n:={\rm arg~min}\,\{d(\u{x})~|~\u{x}\in \u{\kappa}_{n}(\mathcal{B})\},
\end{equation}
in which the function $d:\mathbb{E} \to \mathbb{R}$ is defined by the equation $d(\u{x})=\pr{\u{x}-\lsc[H]{\u{y}}}\cdot \lscH{\u{e}}$.  The value of the function $d$ at  the vector $\u{x}\in \mathbb{E}$ is the signed distance of the point $o+\u{x}$ from ${H}$'s surface.

If $d\pr{\lscH{\u{x}}_n}<0$, then we included  contact interaction for the $n^{\rm th}$ time step. We did this by including the specific contact force measure

$$
\lscH{\t{p}}_n\,\, \delta_{\left(\lscH{\c{X}}_n\right)}\,\, \lscH{\u{a}}
$$ in the specific force measure $\tu{f}_{n}$.
In this last expression
$\lscH{\t{p}}_n$
 is the magnitude of the specific contact force,  $\delta_{\left(\lscH{\c{X}}_n\right)}$ is the Dirac measure located at the material particle $\lscH{\c{X}}_n$, and $\lscH{\u{a}}\in \mathbb{A}$ is defined such that  $\pr{\pr{\lscH{\u{a}}\u{s}}\u{s}}=\lscH{\u{e}}$.
The Dirac measure is defined in~\eqref{eq:DiracMeasureDef}. We model the contact interaction between  $\kappa_{n }\pr{\mathcal{B}}$  and $H$ using the Hertz contact theory.
In accordance with this theory~\cite[Ch. 10]{sackfield2013mechanics}\footnotemark we took

\footnotetext{Equation~\eqref{eq:EllipsoidContactForce} does not explicitly appear in~\cite{sackfield2013mechanics}.  We derived \eqref{eq:EllipsoidContactForce} using some of the results given in ~\cite{sackfield2013mechanics}. Please note the typos in~\cite{sackfield2013mechanics}, especially in  Table 10.2.
}

\begin{equation}
\lscH{\t{p}}_n=\frac{2^{3/2}\pi  }{3 m   }\frac{E}{(1-\nu^2)} \pr{\frac{D(k')}{\kappa_1 K(k')^{3}}}^{1/2} d\left(\lscH{\u{x}}_n\right)^{3/2},
\label{eq:EllipsoidContactForce}
\end{equation}
where, recall that, $m$ is the mass of $\mathcal{B}$, $E$ and $\nu$ are the Young's modulus and the Poisson's ratio of the half space, respectively, $k'$ is either the positive or the negative square root of $1-k^2$, $k\in [-1,1]$
is the solution of the equation
\begin{equation}
k^2\frac{D(\sqrt{1-k^2})}{B(\sqrt{1-k^2})}=\frac{\kappa_1}{\kappa_2},\\
\end{equation}
in which
$D(k):=\pr{K(k)-E(k)}/k^2$ and  $B(k):=K(k)-D(k)$, $\kappa_1$ and $\kappa_2$ are the two principal curvatures of the surface of $\mathcal{B}$ at $\lscH{\c{X}}_n$,   $K(k)$ and $E(k)$ are the values of the complete elliptic integrals of the first and second kind, respectively, at $k$.

\subsection{Extraction of virtual accelerometer data from simulations}
\label{sec:VirtualAccelerometerData}

We took the rigid body in our simulation to be fitted with the number and the type of accelerometers needed for the application of the AO-algorithm. That is, we took it to be fitted with four tri-axial virtual accelerometers. Let the material particles to which the virtual accelerometers were attached be $\lsc{\u{X}}$, where $\mathscr{l}\in \mathcal{J}$, and the orientations of the virtual accelerometers be defined by the orthonormal sets $(\lsc{E}_i)_{i\in \mathcal{I}}$, which  are as defined in \S\ref{Sec:PA}, \eqref{eq:mathscrlEiDef}. Also as defined in \S\ref{Sec:PA}, let $\lsc{\alpha}_{i}(\tau)$  denote the measurement reported by the virtual accelerometer $\lsc{\u{X}}$ for its $i^{\rm th}$ axis, where $i\in \mathcal{I}$, for the time instance $\tau$. We present the procedure that we used for computing  $(\lsc{\alpha}_{i}(n))_{n\in\mathcal{N}}:=(\lsc{\alpha}_{i}(n\Delta \tau))_{n\in\mathcal{N}}$.

From our rigid body simulations, we knew the sequences $(\usf{r}(n))_{n\in \mathcal{N}}$, $(\usf{Q}(n))_{n\in \mathcal{N}}$, $(\usf{W}(n))_{n\in \mathcal{N}}$,  $(\usf{r}''(n))_{n\in \mathcal{N}}$, and $(\usf{W}'(n))_{n\in \mathcal{N}}$. Using the first two of these sequences, we generated
$(\usf{c}(n))_{n\in \mathcal{N}}$ as
\begin{equation}
\usf{c}(n)=\usf{r}(n)-\usf{Q}(n)\usf{r}_{\rm R}.
\label{eq:cnMat}
\end{equation}
Equation~\eqref{eq:cnMat} follows from~\eqref{eq:PosVecMatForm}. Using $(\usf{c}(n))_{n\in \mathcal{N}}$ we then generated the $(\usf{c}''(n))_{n\in \mathcal{N}}$, using the equation

\begin{align}
\boldsymbol{\sf c}''(n) &=
\usf{r}''(n)-\boldsymbol{\sf W}'(n)\left(\usf{r}(n)-\usf{c}(n)\right)-\pr{\usf{W}(n)}^2\left(\usf{r}(n)-\usf{c}(n)\right),
\label{eq:cnppMat}
\end{align}
which follows from~\eqref{eq:SpatialAccFieldMatForm}. Knowing  $\usf{c}''(n)$ we then computed $\lsc{\usf{A}}_n:=\usf{A}_{n \Delta \tau}(\lsc{\usf{X}})$ using the equation
\begin{equation}
\lsc{\usf{A}}_n = \usf{W}'(n)\,\usf{Q}(n)\,\lsc{\usf{X}}+ \pr{\usf{W}(n)}^2\,\,\usf{Q}(n)\,\lsc{\usf{X}} +\boldsymbol{\sf c}''(n),
\end{equation}
which is a consequence of \eqref{eq:PosVecMatForm}, and \eqref{eq:SpatialAccFieldMatForm}. Finally, we construct $(\lsc{\alpha}_{i}(n))_{n\in\mathcal{N}}$ as
\begin{equation}
\lsc{\alpha}_{ i}(n)=
\tps{\lsc{{\usf{A}}}_n}\usf{Q}(n)\, \lsc{\usf{E}}_{i}~\text{(no sum over $\mathscr{l}$)},
\label{eq:alphaSeFbarAseq}
\end{equation}
where $\lsc{\usf{E}}_{i}=\pr{\lsc{\u{E}}_i\cdot \u{E}_j}_{j\in\mathcal{I}}$. Equation~\eqref{eq:alphaSeFbarAseq} follows from \eqref{eq:RelAccAndAccBar}, and \eqref{eq:FicAccAtAcclerometerPoistion}.

\begin{figure}
\centering
\includegraphics[width=\textwidth]{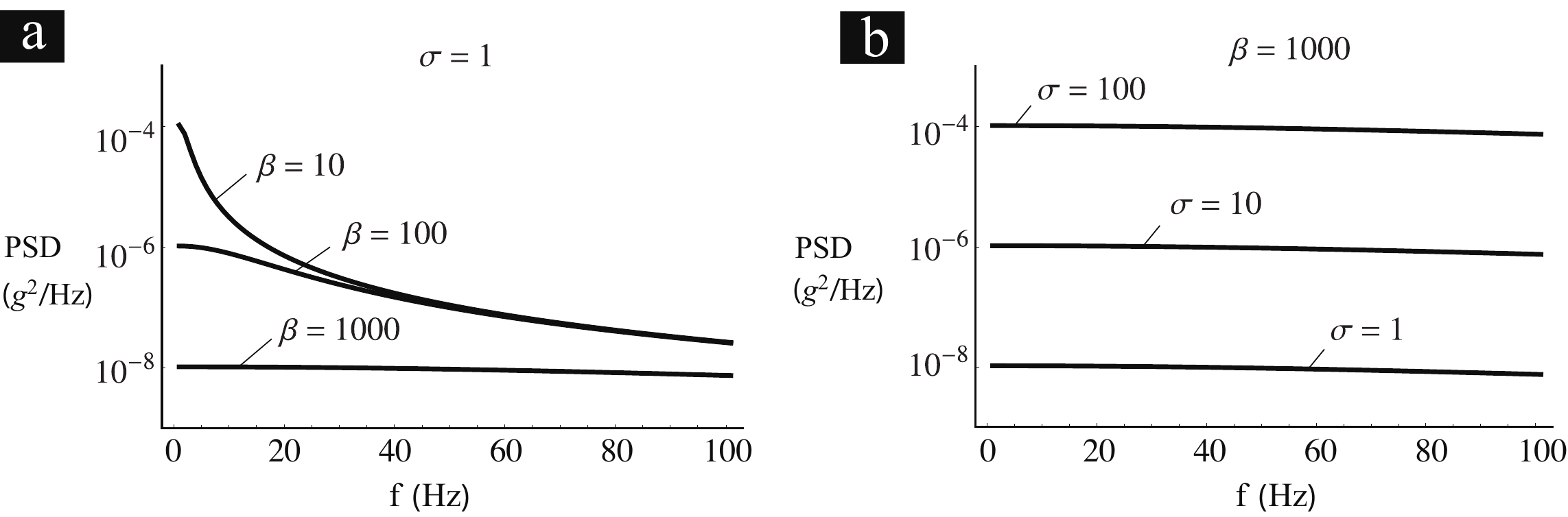}
\caption{Power spectral density (PSD) function of the Ornstein-Uhlenbeck process~\eqref{eq:OUprocess} for different $\sigma$ and $\beta$ values. 
}
\label{fig:PSDplotForDiffSigmaBeta}
\end{figure}

\end{document}